\documentclass[twocolumn]{aastex63}
\newcommand{\rev}{}

\shorttitle{TOI-1749}
\shortauthors{Fukui et al.}
\graphicspath{{./}{figures/}}

\begin{document}

\title{TOI-1749: an M dwarf with a Trio of Planets including a Near-resonant Pair}

\correspondingauthor{Akihiko Fukui}
\email{afukui@g.ecc.u-tokyo.ac.jp}


\author[0000-0002-4909-5763]{A. Fukui}
\affiliation{Komaba Institute for Science, The University of Tokyo, 3-8-1 Komaba, Meguro, Tokyo 153-8902, Japan}
\affiliation{Instituto de Astrof\'isica de Canarias, V\'ia L\'actea s/n, E-38205 La Laguna, Tenerife, Spain}

\author[0000-0002-0076-6239]{J. Korth}
\affiliation{Department of Space, Earth and Environment, Astronomy and Plasma Physics, Chalmers University of Technology, SE-412 96 Gothenburg, Sweden}

\author[0000-0002-4881-3620]{J. H. Livingston}
\affiliation{Department of Astronomy, Graduate School of Science, The University of Tokyo, 7-3-1 Hongo, Bunkyo-ku, Tokyo 113-0033, Japan}

\author[0000-0002-6778-7552]{J. D. Twicken}
\affiliation{SETI Institute, Mountain View, CA  94043, USA}
\affiliation{NASA Ames Research Center, Moffett Field, CA  94035, USA}

\author{M. R. Zapatero Osorio}
\affiliation{Centro de Astrobiolog\'ia (CSIC-INTA), Crta. Ajalvir km 4, E-28850 Torrej\'on de Ardoz, Madrid, Spain}

\author[0000-0002-4715-9460]{J. M. Jenkins}
\affiliation{NASA Ames Research Center, Moffett Field, CA  94035, USA}

\author[0000-0003-1368-6593]{M. Mori}
\affiliation{Department of Astronomy, Graduate School of Science, The University of Tokyo, 7-3-1 Hongo, Bunkyo-ku, Tokyo 113-0033, Japan}

\author{F. Murgas}
\affiliation{Instituto de Astrof\'isica de Canarias, V\'ia L\'actea s/n, E-38205 La Laguna, Tenerife, Spain}
\affiliation{Departamento de Astrof\'isica, Universidad de La Laguna, E-38206 La Laguna, Tenerife, Spain}

\author[0000-0002-8300-7990]{M. Ogihara}
\affiliation{National Astronomical Observatory of Japan, 2-21-1, Osawa, Mitaka, 181-8588 Tokyo, Japan}
\affiliation{Earth-Life Science Institute, Tokyo Institute of Technology, Meguro-ku, Tokyo 152-8550, Japan}

\author[0000-0001-8511-2981]{N. Narita}
\affiliation{Komaba Institute for Science, The University of Tokyo, 3-8-1 Komaba, Meguro, Tokyo 153-8902, Japan}
\affiliation{Japan Science and Technology Agency, PRESTO, 3-8-1 Komaba, Meguro, Tokyo 153-8902, Japan}
\affiliation{Astrobiology Center, 2-21-1 Osawa, Mitaka, Tokyo 181-8588, Japan}
\affiliation{Instituto de Astrof\'isica de Canarias, V\'ia L\'actea s/n, E-38205 La Laguna, Tenerife, Spain}

\author[0000-0003-0987-1593]{E. Pall\'e}
\affiliation{Instituto de Astrof\'isica de Canarias, V\'ia L\'actea s/n, E-38205 La Laguna, Tenerife, Spain}
\affiliation{Departamento de Astrof\'isica, Universidad de La Laguna, E-38206 La Laguna, Tenerife, Spain}

\author{K. G. Stassun}
\affiliation{Department of Physics and Astronomy, Vanderbilt University, 6301 Stevenson Center Lane, Nashville, TN 37235, USA}
\affiliation{Department of Physics, Fisk University, 1000 17th Avenue North, Nashville, TN 37208, USA}

\author[0000-0002-7031-7754]{G. Nowak}
\affiliation{Instituto de Astrof\'isica de Canarias, V\'ia L\'actea s/n, E-38205 La Laguna, Tenerife, Spain}
\affiliation{Departamento de Astrof\'isica, Universidad de La Laguna, E-38206 La Laguna, Tenerife, Spain}

\author{D. R. Ciardi}
\affiliation{Caltech/IPAC-NASA Exoplanet Science Institute, 770 S. Wilson Avenue, Pasadena, CA 91106, USA}


\author{L. Alvarez-Hernandez}
\affiliation{Departamento de Astrof\'isica, Universidad de La Laguna, E-38206 La Laguna, Tenerife, Spain}

\author{V. J. S. B\'ejar}
\affiliation{Instituto de Astrof\'isica de Canarias, V\'ia L\'actea s/n, E-38205 La Laguna, Tenerife, Spain}
\affiliation{Departamento de Astrof\'isica, Universidad de La Laguna, E-38206 La Laguna, Tenerife, Spain}

\author[0000-0002-2891-8222]{N. Casasayas-Barris}
\affiliation{Leiden Observatory, Leiden University, Postbus 9513, 2300 RA Leiden, The Netherlands}

\author[0000-0001-7866-8738]{N. Crouzet}
\affiliation{European Space Agency (ESA), European Space Research and Technology Centre (ESTEC), Keplerlaan 1, 2201 AZ Noordwijk, The Netherlands}

\author[0000-0002-6424-3410]{J. P. de Leon}
\affiliation{Department of Astronomy, Graduate School of Science, The University of Tokyo, 7-3-1 Hongo, Bunkyo-ku, Tokyo 113-0033, Japan}

\author[0000-0002-2341-3233]{E. Esparza-Borges}
\affiliation{Departamento de Astrof\'isica, Universidad de La Laguna, E-38206 La Laguna, Tenerife, Spain}
\affiliation{Departamento de Astrof\'isica, Universidad de La Laguna, E-38206 La Laguna, Tenerife, Spain}

\author{D.~Hidalgo~Soto}
\affiliation{Instituto de Astrof\'isica de Canarias, V\'ia L\'actea s/n, E-38205 La Laguna, Tenerife, Spain}
\affiliation{Departamento de Astrof\'isica, Universidad de La Laguna, E-38206 La Laguna, Tenerife, Spain}

\author[0000-0002-6480-3799]{K. Isogai}
\affiliation{Okayama Observatory, Kyoto University, 3037-5 Honjo, Kamogatacho, Asakuchi, Okayama 719-0232, Japan}
\affiliation{Department of Multi-Disciplinary Sciences, Graduate School of Arts and Sciences, The University of Tokyo, 3-8-1 Komaba, Meguro, Tokyo 153-8902, Japan}

\author[0000-0003-1205-5108]{K. Kawauchi}
\affiliation{Instituto de Astrof\'isica de Canarias, V\'ia L\'actea s/n, E-38205 La Laguna, Tenerife, Spain}
\affiliation{Department of Multi-Disciplinary Sciences, Graduate School of Arts and Sciences, The University of Tokyo, 3-8-1 Komaba, Meguro, Tokyo 153-8902, Japan}

\author{P.~Klagyivik}
\affiliation{Institute of Planetary Research, German Aerospace Center, Rutherfordstrasse 2, D-12489, Berlin, Germany}

\author[0000-0001-9032-5826]{T. Kodama}
\affiliation{Komaba Institute for Science, The University of Tokyo, 3-8-1 Komaba, Meguro, Tokyo 153-8902, Japan}

\author{S. Kurita}
\affiliation{Department of Earth and Planetary Science, Graduate School of Science, The University of Tokyo, 7-3-1 Hongo, Bunkyo-ku, Tokyo 113-0033, Japan}

\author[0000-0001-9194-1268]{N. Kusakabe}
\affiliation{Astrobiology Center, 2-21-1 Osawa, Mitaka, Tokyo 181-8588, Japan}
\affiliation{National Astronomical Observatory of Japan, 2-21-1 Osawa, Mitaka, Tokyo 181-8588, Japan}

\author[0000-0002-4671-2957]{R. Luque}
\affiliation{Instituto de Astrof\'isica de Canarias, V\'ia L\'actea s/n, E-38205 La Laguna, Tenerife, Spain}
\affiliation{Departamento de Astrof\'isica, Universidad de La Laguna, E-38206 La Laguna, Tenerife, Spain}
\affiliation{Instituto de Astrof\'isica de Andaluc\'ia (IAA-CSIC), Glorieta de la Astronom\'ia s/n, E-18008 Granada, Spain}

\author[0000-0002-9510-0893]{A. Madrigal-Aguado}
\affiliation{Instituto de Astrof\'isica de Canarias, V\'ia L\'actea s/n, E-38205 La Laguna, Tenerife, Spain}
\affiliation{Departamento de Astrof\'isica, Universidad de La Laguna, E-38206 La Laguna, Tenerife, Spain}

\author{P. Montanes Rodriguez}
\affiliation{Instituto de Astrof\'isica de Canarias, V\'ia L\'actea s/n, E-38205 La Laguna, Tenerife, Spain}
\affiliation{Departamento de Astrof\'isica, Universidad de La Laguna, E-38206 La Laguna, Tenerife, Spain}

\author{G. Morello}
\affiliation{Instituto de Astrof\'isica de Canarias, V\'ia L\'actea s/n, E-38205 La Laguna, Tenerife, Spain}
\affiliation{Departamento de Astrof\'isica, Universidad de La Laguna, E-38206 La Laguna, Tenerife, Spain}

\author[0000-0003-1510-8981]{T. Nishiumi}
\affiliation{Department of Astronomical Science, The Graduated University for Advanced Studies, SOKENDAI, 2-21-1, Osawa, Mitaka, Tokyo, 181-8588, Japan}
\affiliation{Astrobiology Center, 2-21-1 Osawa, Mitaka, Tokyo 181-8588, Japan}

\author[0000-0003-2066-8959]{J. Orell-Miquel}
\affiliation{Instituto de Astrof\'isica de Canarias, V\'ia L\'actea s/n, E-38205 La Laguna, Tenerife, Spain}
\affiliation{Departamento de Astrof\'isica, Universidad de La Laguna, E-38206 La Laguna, Tenerife, Spain}

\author{M.~Oshagh}
\affiliation{Instituto de Astrof\'isica de Canarias, V\'ia L\'actea s/n, E-38205 La Laguna, Tenerife, Spain}
\affiliation{Departamento de Astrof\'isica, Universidad de La Laguna, E-38206 La Laguna, Tenerife, Spain}

\author[0000-0001-5519-1391]{H. Parviainen}
\affiliation{Instituto de Astrof\'isica de Canarias, V\'ia L\'actea s/n, E-38205 La Laguna, Tenerife, Spain}
\affiliation{Departamento de Astrof\'isica, Universidad de La Laguna, E-38206 La Laguna, Tenerife, Spain}

\author[0000-0003-2693-279X]{M. S\'anchez-Benavente}
\affiliation{Instituto de Astrof\'isica de Canarias, V\'ia L\'actea s/n, E-38205 La Laguna, Tenerife, Spain}
\affiliation{Departamento de Astrof\'isica, Universidad de La Laguna, E-38206 La Laguna, Tenerife, Spain}

\author[0000-0002-1812-8024]{M. Stangret}
\affiliation{Instituto de Astrof\'isica de Canarias, V\'ia L\'actea s/n, E-38205 La Laguna, Tenerife, Spain}
\affiliation{Departamento de Astrof\'isica, Universidad de La Laguna, E-38206 La Laguna, Tenerife, Spain}

\author[0000-0003-2887-6381]{Y. Terada}
\affiliation{Institute of Astronomy and Astrophysics, Academia Sinica, P.O. Box 23-141, Taipei 10617, Taiwan, R.O.C.}
\affiliation{Department of Astrophysics, National Taiwan University, Taipei 10617, Taiwan, R.O.C.}

\author[0000-0002-7522-8195]{N. Watanabe}
\affiliation{Department of Multi-Disciplinary Sciences, Graduate School of Arts and Sciences, The University of Tokyo, 3-8-1 Komaba, Meguro, Tokyo 153-8902, Japan}

\author{G. Chen}
\affiliation{Key Laboratory of Planetary Sciences, Purple Mountain Observatory, Chinese Academy of Sciences, Nanjing
210023, PR China}

\author{M. Tamura}
\affiliation{Department of Astronomy, Graduate School of Science, The University of Tokyo, 7-3-1 Hongo, Bunkyo-ku, Tokyo 113-0033, Japan}
\affiliation{Astrobiology Center, 2-21-1 Osawa, Mitaka, Tokyo 181-8588, Japan}
\affiliation{National Astronomical Observatory of Japan, 2-21-1 Osawa, Mitaka, Tokyo 181-8588, Japan}


\author{P. Bosch-Cabot}
\affiliation{Observatori Astron\`{o}mic Albany\`{a}, Cam\'{i} de Bassegoda S/N, Albany\`{a} E-17733, Girona, Spain}

\author{M. Bowen}
\affiliation{Department of Physics and Astronomy, George Mason University, Fairfax, VA, USA}

\author{K. Eastridge}
\affiliation{Department of Physics and Astronomy, George Mason University, Fairfax, VA, USA}

\author{L. Freour}
\affiliation{DTU Space, National Space Institute, Technical University of Denmark, Elektrovej 328, DK-2800 Kgs. Lyngby, Denmark}

\author{E. Gonzales}
\affiliation{National Science Foundation Graduate Research Fellow.; University of California, Santa Cruz, 1156 High Street, Santa Cruz, CA 95064, USA}

\author{P. Guerra}
\affiliation{Observatori Astron\`{o}mic Albany\`{a}, Cam\'{i} de Bassegoda S/N, Albany\`{a} E-17733, Girona, Spain}

\author{Y. Jundiyeh}
\affiliation{DTU Space, National Space Institute, Technical University of Denmark, Elektrovej 328, DK-2800 Kgs. Lyngby, Denmark}

\author{T. K. Kim}
\affiliation{Department of Physics and Astronomy, George Mason University, Fairfax, VA, USA}
\affiliation{Thomas Jefferson High School for Science and Technology, 6560 Braddock Road, Alexandria, VA 22312, USA}

\author{L. V. Kroer}
\affiliation{DTU Space, National Space Institute, Technical University of Denmark, Elektrovej 328, DK-2800 Kgs. Lyngby, Denmark}

\author[0000-0001-8172-0453]{A.~M.~Levine}
\affiliation{Department of Physics and Kavli Institute for Astrophysics and Space Research, Massachusetts Institute of Technology, 77 Massachusetts Avenue, Cambridge, MA 02139, USA}

\author{E.~H.~Morgan} 
\affiliation{Department of Physics and Kavli Institute for Astrophysics and Space Research, Massachusetts Institute of Technology, 77 Massachusetts Avenue, Cambridge, MA 02139, USA}

\author[ 0000-0003-4701-8497]{M. Reefe}
\affiliation{Department of Physics and Astronomy, George Mason University, Fairfax, VA, USA}

\author[0000-0003-1001-0707]{R. Tronsgaard}
\affiliation{DTU Space, National Space Institute, Technical University of Denmark, Elektrovej 328, DK-2800 Kgs. Lyngby, Denmark}

\author{C. K. Wedderkopp}
\affiliation{DTU Space, National Space Institute, Technical University of Denmark, Elektrovej 328, DK-2800 Kgs. Lyngby, Denmark}

\author{J. Wittrock}
\affiliation{Department of Physics and Astronomy, George Mason University, Fairfax, VA, USA}


\author[0000-0001-6588-9574]{K. A. Collins}
\affiliation{Center for Astrophysics \textbar \ Harvard \& Smithsonian, 60 Garden Street, Cambridge, MA 02138, USA}

\author{K. Hesse}
\affiliation{Department of Astronomy, Wesleyan University, Middletown, CT 06459, USA}


\author[0000-0003-2058-6662]{D. W. Latham}
\affiliation{Center for Astrophysics \textbar \ Harvard \& Smithsonian, 60 Garden Street, Cambridge, MA 02138, USA}

\author{G. R. Ricker}
\affiliation{Department of Physics and Kavli Institute for Astrophysics and Space Research, Massachusetts Institute of Technology, 77 Massachusetts Avenue, Cambridge, MA 02139, USA}

\author[0000-0002-6892-6948]{S.~Seager}
\affiliation{Department of Physics and Kavli Institute for Astrophysics and Space Research, Massachusetts Institute of Technology, 77 Massachusetts Avenue, Cambridge, MA 02139, USA}
\affiliation{Department of Earth, Atmospheric and Planetary Sciences, Massachusetts Institute of Technology, 77 Massachusetts Avenue, Cambridge, MA 02139, USA}
\affiliation{Department of Aeronautics and Astronautics, Massachusetts Institute of Technology, 77 Massachusetts Avenue, Cambridge, MA 02139, USA}

\author{R. Vanderspek}
\affiliation{Department of Physics and Kavli Institute for Astrophysics and Space Research, Massachusetts Institute of Technology, 77 Massachusetts Avenue, Cambridge, MA 02139, USA}

\author{J. Winn}
\affiliation{Department of Astrophysical Sciences, Princeton University, 4 Ivy Lane, Princeton, NJ 08540, USA}


\author{E. Bachelet}
\affiliation{Las Cumbres Observatory, 6740 Cortona Drive, Suite 102, Goleta, CA 93117-5575, USA}

\author[0000-0002-9716-6175]{M. Bowman}
\affiliation{Las Cumbres Observatory, 6740 Cortona Drive, Suite 102, Goleta, CA 93117-5575, USA}

\author{C. McCully}
\affiliation{Las Cumbres Observatory, 6740 Cortona Drive, Suite 102, Goleta, CA 93117-5575, USA}

\author{M. Daily}
\affiliation{Las Cumbres Observatory, 6740 Cortona Drive, Suite 102, Goleta, CA 93117-5575, USA}

\author[0000-0002-8590-007X]{D. Harbeck}
\affiliation{Las Cumbres Observatory, 6740 Cortona Drive, Suite 102, Goleta, CA 93117-5575, USA}

\author{N. H. Volgenau}
\affiliation{Las Cumbres Observatory, 6740 Cortona Drive, Suite 102, Goleta, CA 93117-5575, USA}





\begin{abstract}

We report the discovery of one super-Earth\rev{-} (TOI-1749b) and two sub-Neptune\rev{-sized planets} (TOI-1749c and TOI-1749d) transiting an early M dwarf at a distance of 100~pc, which were first identified as planetary candidates using data from the TESS photometric survey. 
We have followed up this system from the ground by means of multiband transit photometry, adaptive optics imaging, and low-resolution spectroscopy, from which we have validated the planetary nature of the candidates. We find that TOI-1749b, c, and d have orbital periods of 2.39, 4.49, and 9.05 days, and radii of 1.4, 2.1, and 2.5 $R_\oplus$, respectively. We also place 95\% confidence upper limits on the masses of 57, 14, and 15 $M_\oplus$ for TOI-1749b, c, and d, respectively, from transit timing variations. The periods, sizes, and tentative masses of these planets are in line with a scenario in which all three planets initially had a hydrogen envelope on top of a rocky core, and only the envelope of the innermost planet has been stripped away by photoevaporation and/or core-powered mass-loss mechanisms. These planets are similar to other planetary trios found around M dwarfs, such as TOI-175b,c,d and TOI-270b,c,d, in the sense that the outer pair has a period ratio within 1\% of 2. Such a characteristic orbital configuration, in which an additional planet is located interior to a near 2:1 period-ratio pair, is relatively rare around FGK dwarfs.

\end{abstract}



\section{Introduction} 
\label{sec:intro}

Kepler has revealed that super-Earth- and sub-Neptune-sized planets are common in close orbits \citep{2011ApJ...736...19B,2012ApJS..201...15H}.
Among these planetary systems, those hosting multiple planets are a key to understanding how they have formed and evolved. In particular, planets in or near mean motion resonances (MMRs) hold valuable information about their formation and evolution histories, because they have probably been captured in an MMR as a consequence of convergent migration \citep[e.g.,][]{2002ApJ...567..596L}. It has been revealed that planetary pairs close to MMRs are not the dominant population in the Kepler multis; instead, there is a significant excess in orbital period ratio of adjacent pairs just outside of exact commensurabilities \citep{2011ApJS..197....8L,2014ApJ...790..146F,2015MNRAS.448.1956S}. So far several mechanisms have been proposed to explain the observed period-ratio distribution \citep[e.g.,][see Section \ref{sec:period_ratio}]{2012ApJ...756L..11L,2013ApJ...778....7B,2013AJ....145....1B,2015ApJ...803...33C,2017MNRAS.470.1750I}, although the dominant mechanisms that produce the distribution have not yet been fully understood.

Transiting planets in or near MMRs are also valuable to measure their masses through transit timing variations (TTVs), because the amplitude of TTV signals is amplified when a planetary pair is close to an MMR \citep{2005MNRAS.359..567A,2005Sci...307.1288H}. The TTV amplitudes are also proportional to the orbital periods of the planets, which makes this technique advantageous over the radial velocity technique to measure the masses of longer period planets \citep{2016MNRAS.457.4384S}; the radial velocity amplitude varies as $P^{-1/3}$, where $P$ is the orbital period of the planet. Measuring the masses and radii of long-period planets near MMRs is of particular importance in understanding how much gas a protoplanetary core can accumulate from the surrounding disk nebula. This is because they may retain primordial envelopes without significant atmospheric loss due to strong stellar irradiation or giant impacts, the latter of which would break the near-MMR orbits \citep{2017MNRAS.470.1750I}.

Although Kepler has discovered a number of planetary pairs close to MMRs around FGK dwarfs, the number of such pairs has still been limited around M dwarfs.
The successor of Kepler, the Transiting Exoplanet Survey Satellite \citep[TESS,][]{2015JATIS...1a4003R}, has been increasing the number of known planetary pairs close to MMRs \rev{including those around M dwarfs, which can be important samples to study the dependence of the properties (such as period ratio, radius, and mass) of near-MMR planets on stellar mass.}
However, because \rev{near-MMR planetary pairs} are relatively rare as revealed by Kepler \citep[only $\sim$16\% of Kepler's multiplanetary systems contain planetary pairs close to 2:1 commensurability;][]{2011ApJS..197....8L}, those around bright host stars that allow precise radial velocity (RV) measurements are limited. In addition, in many cases the time span of the TESS observations is not long enough to cover the timescale of a TTV signal, and/or the photometric precision of TESS is not high enough to measure the times of individual transits with a sufficient precision, making mass measurements by the TTV method challenging with the TESS data alone. 

Following up TTVs from the ground is therefore an important and complementary way to measure the masses of near-MMR planets discovered by TESS \citep[e.g.,][]{2020A&A...642A..49D,2021AJ....161..161D}. For this purpose, multiband simultaneous cameras mounted on 1--2m class telescopes, such as the MuSCAT series \citep{2015JATIS...1d5001N,2019JATIS...5a5001N,2020SPIE11447E..5KN}, are especially efficient; in addition to the fact that a lot of telescope time is available for this class of telescopes compared to larger ones, the redundant channels in a multiband camera can improve the precision of transit timing measurements \citep[e.g.,][]{2015ApJ...815...47N} and also ensure the robust detection of shallow ($<$ 0.1\%) transit signals even from the ground \citep{2016AJ....152..171F}.

Here, we report the detection and follow-up observations of a multiplanetary system hosted by a relatively faint M dwarf, TOI-1749, in which two sub-Neptune-sized planets reside very close to the 2:1 commensurability.
While mass measurements of these planets are challenging for most of the current RV facilities due to the faintness of the host star, from ground-based photometric observations with various instruments including the multiband imagers MuSCAT2 and MuSCAT3, we have succeeded in constraining the masses of the planets solely by TTVs.

This paper is organized as follows. In Section \ref{sec:observations}, we describe the TESS photometric data and ground-based follow-up observations of the TOI-1749 system. In Section \ref{sec:analysis}, we provide data analyses including stellar characterization, signal search for a third transiting planet, transit model fitting to the TESS and ground-based light curves, validation of the planetary candidates, and photodynamical modeling. We discuss the characteristics of this system in Section \ref{sec:discussion}, and summarize the paper in Section \ref{sec:summary}.

\section{Observations}
\label{sec:observations}

\subsection{TESS Photometry}

TOI-1749 (TIC~233602827) was observed by TESS with a 2 minutes cadence in 12 TESS sectors, specifically, Sectors 14--21 (from 2019 July 18 to 2020 February 18) and Sectors 23--26 (from 2020 March 18 to 2020 July 4). The coordinates and magnitudes of the target star are summarized in Table \ref{tab:stellar_properties}.
The collected data were processed with a pipeline developed by the TESS Science Processing Operations Center (SPOC) at NASA Ames Research Center \citep{2016SPIE.9913E..3EJ}, from which two transit signals with orbital periods of 4.49 and 9.05 days were detected \citep{2002ApJ...575..493J,2010SPIE.7740E..0DJ,2019PASP..131b4506L} and were labeled as TOI-1749.01 and TOI-1749.02, respectively \citep{2021ApJS..254...39G}. In addition, 
we detected an additional transiting planetary candidate interior to these two candidates as we describe in Section \ref{sec:TESS_03}, which we internally labeled as TOI-1749.03. As we describe in Section \ref{sec:validation}, we validate these three planetary candidates as bona fide planets. Hereafter we designate TOI-1749.03, TOI-1749.01, and TOI-1749.02 as TOI-1749b, TOI-1749c, and TOI-1749d, respectively, in order of orbital period.

\begin{figure}[htb!]
\centering
\includegraphics[width=8.6cm]{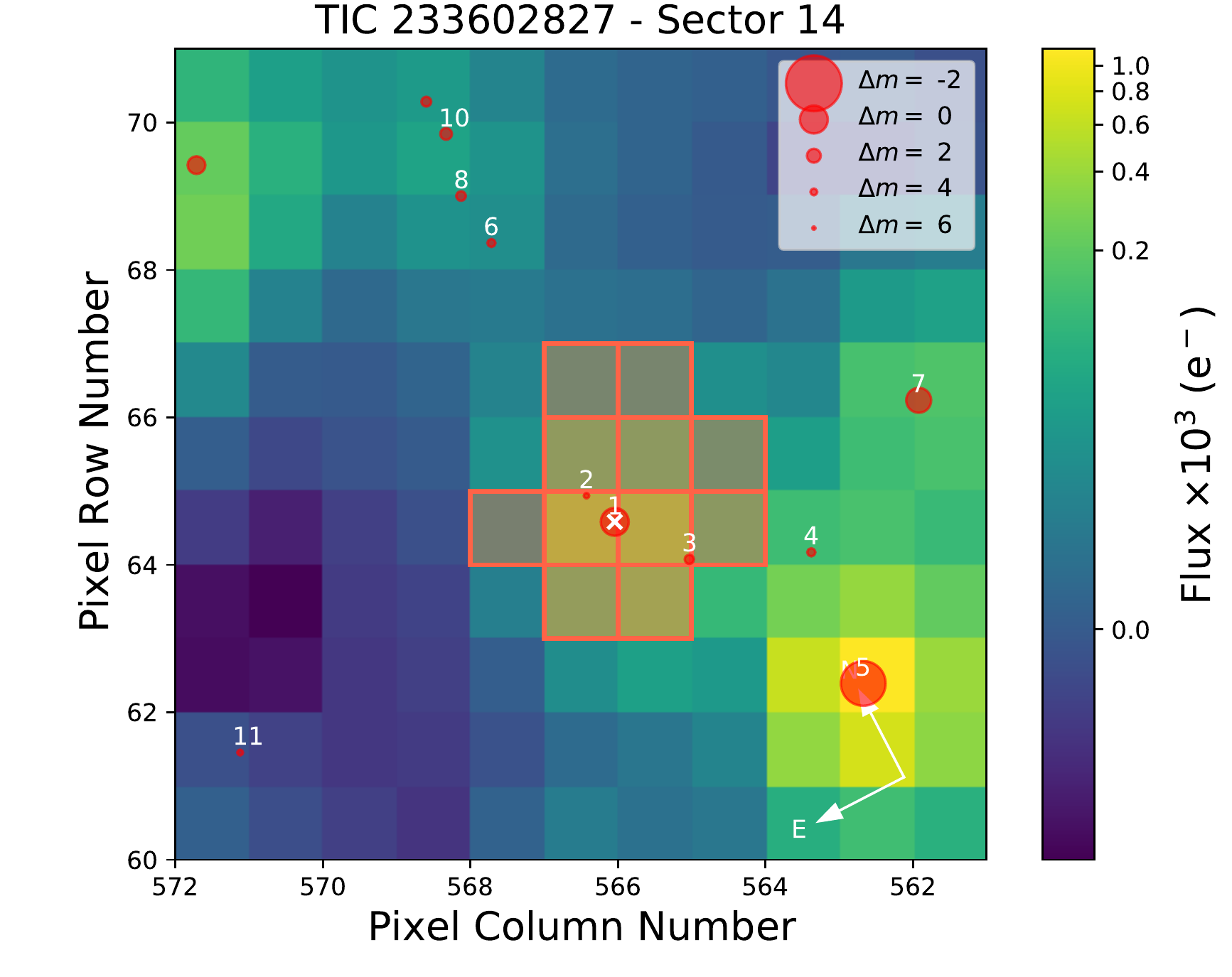}
\caption{Target pixel file (TPF) data around TOI-1749 (marked by white cross) from Sector 14 created with {\tt tpfplotter} \citep{2020A&A...635A.128A}. The red circles are the sources in the Gaia DR2 catalog, where the sizes are scaled by magnitude. The orange-outlined pixels constitute the aperture mask used to extract the photometry. \label{fig:TPF}}
\end{figure}

For further light-curve analyses, we extracted the Presearch Data Conditioning Simple Aperture Photometry (PDC-SAP) \citep{2012PASP..124.1000S,2012PASP..124..985S,2014PASP..126..100S} from the Mikulski Archive for Space Telescopes (MAST) at the Space Telescope Science Institute (STScI).  Figure \ref{fig:TPF} shows an example of the TESS target pixel file (TPF) data around the target star, in which the aperture mask used by the pipeline to extract the light curve is indicated by orange-outlined pixels. There are two other sources of the Gaia DR2 catalog in the aperture masks (labeled as 2 and 3 in Figure \ref{fig:TPF}), the fluxes of which were subtracted from the PDC-SAP fluxes, which are in absolute flux scale. We removed data points that are flagged as suspect in regard to quality from the PDC-SAP light curve, and then normalized it so that the median flux value in each sector is unity.

\begin{deluxetable}{lcc}
\label{tab:stellar_properties}
\tablecaption{Properties of the Host Star}
\tablehead{
\colhead{Parameter} & \colhead{Value} & \colhead{Sources}
}
\startdata
\multicolumn{3}{c}{\it{Stellar name}}\\
TOI & 1749 & (1)\\
TIC & 233602827 & (2)\\
\multicolumn{3}{c}{\it{Astrometric and kinematic information}}\\
$\alpha$ & 18:50:56.93 & (2)\\
$\delta$ & 64:25:10.08 & (2)\\
Distance (pc) & $99.56 \pm 0.12$ & (2)\\
Parallax (mas) & $10.0582 \pm 0.0095$ & (3)\\
$\mu_{\alpha}$\,cos\,$\delta$ (mas\,yr$^{-1}$)  & $-54.347 \pm 0.012$ & (3)\\
$\mu_{\delta}$ (mas\,yr$^{-1}$) & $+61.844 \pm 0.012$ & (3)\\
$U$ (km s$^{-1}$) & $-21.38 \pm 0.17$  & This work\\
$V$ (km s$^{-1}$) & $-16.44 \pm 2.02$ & This work\\
$W$ (km s$^{-1}$) & $27.96 \pm 0.92$  & This work\\
\multicolumn{3}{c}{\it{Magnitudes}}\\
TESS & $12.2574 \pm 0.0073$ & (2)\\
$G$ & $13.1798 \pm 0.0003$ & (3)\\
$V$ & $13.86 \pm 0.072$ & (2)\\
$J$ & $11.069 \pm 0.023$ & (4)\\
$H$ & $10.446 \pm 0.021$ & (4)\\
$K$ & $10.270 \pm 0.019$ & (4)\\
\multicolumn{3}{c}{\it{Physical parameters}}\\
Sp. type & M0V $\pm$ 0.5 subtype & This work\\
Mass ($M_\odot$) & $0.58 \pm 0.03$ & This work\\
Radius ($R_\odot$) & $0.55 \pm 0.03$ & This work\\
$T_{\rm eff}$ (K) & $3985 \pm 55$ & This work\\
$\log g$ (cgs) & $4.70 \pm 0.05$ & This work\\
$[$Fe/H$]$ (dex) & $-0.26 \pm 0.08$ & This work\\
Age (Gyr) & $>$0.8 & This work\\
\enddata
\tablecomments{Sources: (1) \citet{2021ApJS..254...39G}, (2) TIC v8 \citep{Stassun_2019}, 
(3) Gaia EDR3 \citep{2021AA...649A...1G}, 
(4) 2MASS \citep{2006AJ....131.1163S}}
\end{deluxetable}

\subsection{Ground-based Transit Observations}
\label{sec:GR_transit_obs}

To confirm the transit signals detected by TESS and measure TTVs in the system, we conducted photometric (or spectrophotometric) observations of predicted transits of the TOI-1749 planets using various ground-based telescopes. The telescopes and instruments that we used for these observations are listed in Table \ref{tbl:list_of_instruments}, and the observed transits are summarized in Table \ref{tbl:observing_log}. In the following subsections, we briefly describe observations and data reductions for data that are used in the subsequent analyses. Other observations that are not used in the subsequent analyses are described in the Appendix.

\subsubsection{TCS 1.52m / MuSCAT2}

We observed one, five, and four transits of TOI-1749b, TOI-1749c, and TOI-1749d, respectively, between 2020 June and 2020 October with the multiband imager MuSCAT2 \citep{2019JATIS...5a5001N}, which is mounted on the 1.52~m TCS telescope of the Teide Observatory at Tenerife in the Canary Islands (Spain). 
MuSCAT2 has four optical channels, each installed with a 1024$\times$1024 pixels CCD camera, enabling $g$-, $r$-, $i$-, and $z_s$-band simultaneous imaging. Each camera has a pixel scale of 0\farcs44 pixel$^{-1}$, providing a field of view (FOV) of 7\farcm4 $\times$ 7\farcm4. The exposure times were set to values in the range 15--30 s depending on the channel and sky conditions. On 2020 September 1 UT, the $g$-band channel was not available and the observation was conducted with the remaining three channels.

We applied dark-image and flat-field corrections to the acquired images, and then performed aperture photometry on the calibrated images for the target and several comparison stars using a custom pipeline \citep{2011PASJ...63..287F}. The combination of comparison stars and aperture size was optimized for each dataset (for each night and each band) so that the dispersion of the produced light curve, after removing apparent outliers, was minimized. The adopted aperture radii range from 6 to 12 pixels, or 2\farcs6--5\farcs2, which are not significantly contaminated by any of the Gaia sources seen in Figure \ref{fig:TPF}. We converted all time stamps of the light curves into Barycentric Julian Dates in Barycentric Dynamical Time (BJD$_{\rm TDB}$) using the converter of \citet{2010PASP..122..935E}.

\subsubsection{LCO 1m / Sinistro}

We observed one partial transit (ingress) of TOI-1749d on 2020 September 3 UT with the single-band imager Sinistro mounted on one of the 1~m telescopes of Las Cumbres Observatory (LCO) at McDonald Observatory (USA). Sinistro is equipped with a 4k $\times$ 4k CCD with a pixel scale of 0\farcs389 pixel$^{-1}$, providing an FOV of 26\farcm5 $\times$ 26\farcm5. We observed the target field in $i$ band with the exposure time of 50~s in the full-frame mode, with the image slightly out of focus.
The obtained raw images were processed by the {\tt BANZAI} pipeline \citep{curtis_mccully_2018_1257560} to perform dark-image and flat-field corrections, and then aperture photometry was performed in the same way as for the MuSCAT2 data.

\subsubsection{LCO 2m / Spectral}

We observed a partial transit (egress) and a full transit of TOI-1749d on 2020 August 25 UT and 2020 September 12 UT, respectively, with the single-band imager Spectral, which was mounted on the 2m Faulkes Telescope North (FTN) of LCO at Haleakala Observatory in Hawaii. Spectral was equipped with a 4k $\times$ 4k CCD with a pixel scale of 0\farcs152 pixel$^{-1}$, providing an FOV of 10\farcm5 $\times$ 10\farcm5. All of the observations were done in $i$ band with the 2x2 binning mode. The exposure times were set to values in the range 20-30~s depending on the night, and the detector was slightly out of focus.
The observed images were reduced in the same way as the Sinistro data.

\subsubsection{LCO 2m / MuSCAT3}

In 2020 September, the new multiband imager MuSCAT3 was installed on the FTN as a replacement for Spectral \citep{2020SPIE11447E..5KN}. 
As with MuSCAT2, MuSCAT3 has four channels for the $g$, $r$, $i$, and $z_s$ bands, but has wider format CCD cameras with a pixel array size of 2k $\times$ 2k.
The pixel scale of each camera is 0\farcs266 pixel$^{-1}$, which provides an FOV of 9\farcm1 $\times$ 9\farcm1. One partial transit (ingress) of TOI-1749c was observed with MuSCAT3 on 2020 October 21 UT, during the commissioning phase of the instrument. The observation was done slightly out of focus with exposure times of 45, 15, 25, and 45~s for the $g$, $r$, $i$, and $z_s$ bands, respectively.
The observed images were reduced in the same way as the Sinistro data.

\begin{deluxetable*}{lccccc}[htp]
\label{tbl:list_of_instruments}
\tablecaption{\rev{List of telescopes and instruments used for ground-based transit observations.}}
\tablehead{
 \colhead{Telescope/Instrument} & \colhead{Aperture} & \colhead{Observatory} &  \colhead{Pixel scale} & \colhead{FoV} & \# of imaging channels\\
& (m) & & (pixel$^{-1}$) & (or slit size) & (or spectral resolution)
}
\startdata
TCS/MuSCAT2  & 1.52 & Teide &  0\farcs44 & 7\farcm4 $\times$ 7\farcm4 & 4\\
LCO 1m/Sinistro  & 1.0 & McDonald &  0\farcs389  & 26\farcm5$\times$ 26\farcm5 & 1\\
LCO 2m/Spectral & 2.0 & Haleakala &  0\farcs304 & 10\farcm5 $\times$ 10\farcm5 & 1\\[-3pt]
&&& (2x2 binned) &\\
LCO 2m/MuSCAT3  & 2.0 & Haleakala &  0\farcs266 & 9\farcm1 $\times$ 9\farcm1 & 4\\
GMU 0.81m & 0.81 & GMU & 0\farcs34 & $23' \times 23'$ & 1\\
NOT/ALFOSC & 2.56 & Roque de los Muchachos & 0\farcs214 & 7\farcm3 $\times$ 7\farcm3 & 1\\
OAA 0.4m & 0.4 & OAA & 1\farcs44 & $36' \times 36'$ & 1\\ 
GTC/OSIRIS & 10 & Roque de los Muchachos & 0\farcs254 & ($40'' \times 7\farcm4$) & (1122)\\[-3pt]
&&& (2x2 binned) &\\
\enddata
\end{deluxetable*}

\begin{deluxetable*}{lcccc}[htp]
\label{tbl:observing_log}
\tablecaption{Summary of photometric transit observations from the ground.}
\tablehead{
\colhead{Date (UT)} & \colhead{Telescope/Instrument} & \colhead{Filters} & \colhead{Transit Coverage} &
\colhead{Global Fit}\\
& & (or grism) & &
}
\startdata
\multicolumn{5}{c}{TOI-1749b}\\
2020 Oct. 10 & TCS/MuSCAT2 & $g$, $r$, $i$, $z_s$ & {\rm egress} & \checkmark\\
2020 Oct. 10 & GTC/OSIRIS &  (R1000R) & {\rm full} & \\
\hline
\multicolumn{5}{c}{TOI-1749c}\\
2020 Apr. 11 &  GMU 0.81m & $R$ & {\rm full} &\\
2020 Jul. 9 & TCS/MuSCAT2 & $g$, $r$, $i$, $z_s$ & {\rm full} & \checkmark\\
2020 Aug. 14 & TCS/MuSCAT2 & $g$, $r$, $i$, $z_s$ & {\rm full} & \checkmark\\
2020 Aug. 14 & NOT/ALFOSC & $R$ & {\rm egress} &\\
2020 Aug. 23 & TCS/MuSCAT2 & $g$, $r$, $i$, $z_s$ & {\rm egress} & \checkmark\\
2020 Sep. 1 & TCS/MuSCAT2 & $r$, $i$, $z_s$ & {\rm full} & \checkmark\\
2020 Oct. 7 & TCS/MuSCAT2 & $g$, $r$, $i$, $z_s$ & {\rm egress} & \checkmark\\
2020 Oct. 21 & LCO 2m/MuSCAT3 & $g$, $r$, $i$, $z_s$ & {\rm ingress} & \checkmark\\
\hline
\multicolumn{5}{c}{TOI-1749d}\\
2020 Jun. 22 & TCS/MuSCAT2 & $g$, $r$, $i$, $z_s$ & {\rm egress} & \checkmark\\
2020 Jul. 1 & TCS/MuSCAT2 & $g$, $r$, $i$, $z_s$ & {\rm full} & \checkmark\\
2020 Jul. 10 & TCS/MuSCAT2 & $g$, $r$, $i$, $z_s$ & {\rm full} & \checkmark\\
2020 Jul. 10 & OAA 0.4m & $I$ & {\rm full} &\\
2020 Jul. 19 & TCS/MuSCAT2 & $g$, $r$, $i$, $z_s$ & {\rm full} & \checkmark\\
2020 Aug. 25 & LCO 2m/Spectral & $i$ & {\rm egress} & \checkmark\\
2020 Sep. 3 & LCO 1m/Sinistro & $i$ & {\rm ingress} & \checkmark\\
2020 Sep. 12 & LCO 2m/Spectral & $i$ & {\rm full} & \checkmark\\
\enddata
\end{deluxetable*}

\subsection{Low-resolution Spectroscopy with NOT/ALFOSC}
\label{subsec-not_alfosc}
\begin{figure}[htp]
    \centering
    \plotone{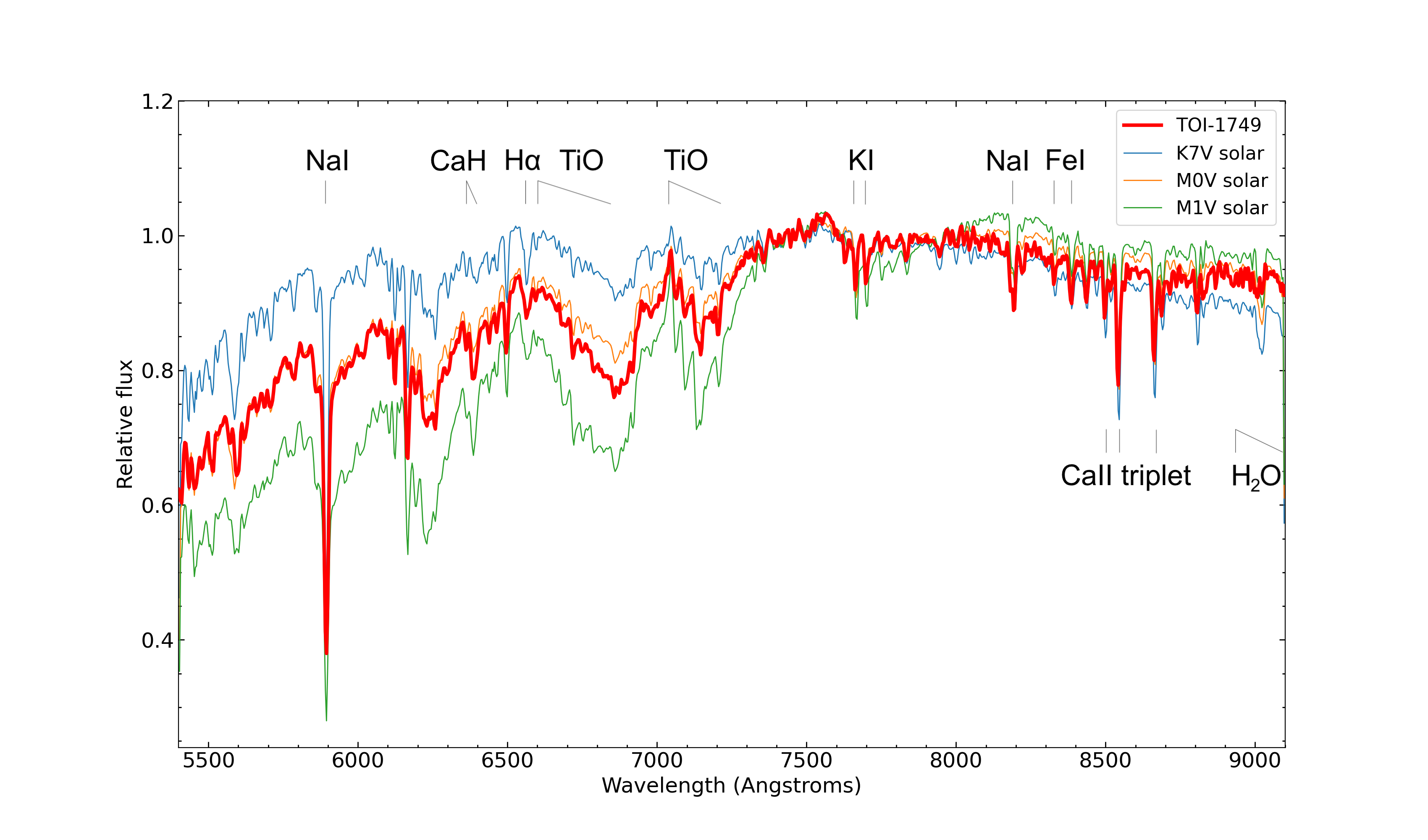}
    \caption{ALFOSC optical, low-resolution spectrum of TOI-1749 (red line, $R \sim 450$ at 710 nm). The observed spectrum is corrected for telluric absorption. Three reference K7-, M0-, and M1-type spectra with solar metallicity from the database of \cite{2017ApJS..230...16K} are also shown for comparison. All data are normalized to unity at around 750 nm. Some of the strongest atomic and molecular absorption features are labeled.
    \label{TOI-1749-spectrum-not_alfosc}}
\end{figure}

On 2020 September 5 UT, we obtained the optical low-resolution spectrum of TOI-1749 with \rev{the Alhambra Faint Object Spectrograph and Camera (ALFOSC) on the 2.56~m Nordic Optical Telescope (NOT) at Roque de los Muchachos Observatory on La Palma in the Canary Islands} under the observing program 59-210. ALFOSC is equipped with a 2048$\times$2064 CCD detector with a pixel scale of 0\farcs2138 pixel$^{-1}$.
We used grism number 5 and a horizontal long slit with a width of 1\farcs0, which yield a nominal spectral dispersion of 3.53 \AA~pixel$^{-1}$ and a usable wavelength space coverage between 5000 and 9100~\AA. The ALFOSC spectrum was acquired with an exposure time of 1800~s at the parallactic angle and an airmass of 1.26. On the same night, we acquired an ALFOSC spectrum of the spectrophotometric standard star G191--B2B (white dwarf) with the same instrumental setup as TOI-1749, at an airmass of 1.18 and exposure time of 180 s. Raw ALFOSC images were reduced following standard procedures at optical wavelengths: bias subtraction, flat-fielding using spectral halogen flat fields, and optimal extraction using appropriate packages within the IRAF\footnote{Image Reduction and Analysis Facility (IRAF) is distributed by the National Optical Astronomy Observatories, which are operated by the Association of Universities for Research in Astronomy, Inc., under contract with the National Science Foundation.} environment. Wavelength calibration was performed with a precision of 0.65 \AA~using He\,{\sc i} and Ne\,{\sc i} arc lines observed on the same night. The instrumental response was corrected using observations of the spectrophotometric standard star G191--B2B. Because the primary target and the standard star were observed close in time and at a similar airmass, we corrected for telluric lines absorption by dividing the target spectrum by the spectrum of the standard normalized to the continuum. The final low-resolution spectrum of TOI-1749 depicted in Figure~\ref{TOI-1749-spectrum-not_alfosc}
has a spectral resolution of 16 \AA{} ($R \approx 450$ at 7100~\AA{}).

\subsection{Adaptive Optics Imaging}

As part of our standard process for validating transiting exoplanets to assess the possible contamination of bound or unbound companions on the derived planetary radii \citep{ciardi2015}, TOI~1749 was observed with infrared high-resolution adaptive optics (AO) imaging at the Keck Observatory. The observations were made with the NIRC2 instrument on Keck II behind the natural guide star AO system \citep{wizinowich2000} on 2020~May~28 UT in the standard 3-point dither pattern that is used with NIRC2 to avoid the left lower quadrant of the detector, which is typically noisier than the other three quadrants. The dither pattern step size was $3\arcsec$ and was repeated four times.  

The observations were made in the narrowband $Br-\gamma$ filter $(\lambda_o = 2.1686; \Delta\lambda = 0.0326 \mu$m) with an integration time of 1~s with one coadd per frame for a total of 3~s on target.  The camera was in the narrow-angle mode with a full FOV of $\sim10\arcsec$ and a pixel scale of 0\farcs099442 per pixel.  The AO data were processed and analyzed with a custom set of IDL tools.  The science frames were flat-fielded and sky-subtracted.  The flat fields were generated from a median average of dark-subtracted flats taken on-sky.  The flats were normalized such that the median value of the flats is unity.  The sky frames were generated from the median average of the 15 dithered science frames; each science frame was then sky-subtracted and flat-fielded.  The reduced science frames were combined into a single combined image using a intra-pixel interpolation that conserves flux, shifts the individual dithered frames by the appropriate fractional pixels, and median-coadds the frames.  The final resolution of the combined dithers was 0\farcs050, which was determined from the FWHM of the point spread function (PSF).

\begin{figure}[htp]
    \centering
    \includegraphics[width=8cm]{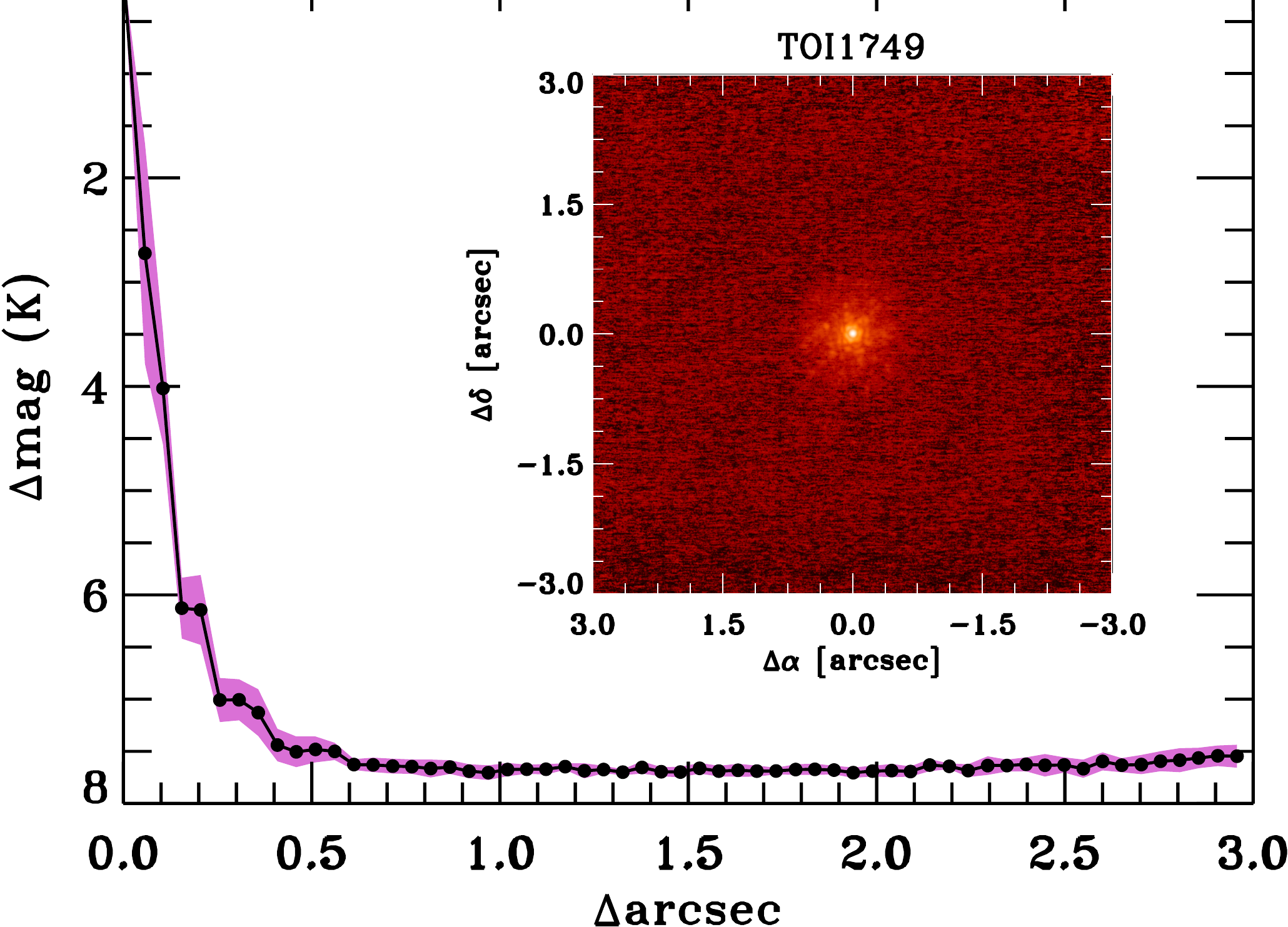}
    \caption{Companion sensitivity for the Keck adaptive optics imaging.  The black points represent the 5$\sigma$ limits and are separated in steps of 1 FWHM ($\sim$0\farcs051); the purple band represents the azimuthal dispersion ($\pm$1$\sigma$) of the contrast determinations (see text). The inset image shows no stars in addition to the primary target.}
    \label{fig:AO}
\end{figure}

The Keck AO observations revealed no additional stellar companions within a resolution of $\sim$0\farcs051 FWHM (Figure \ref{fig:AO}). The sensitivities of the final combined AO image were determined by injecting simulated sources azimuthally around the primary target every $20^\circ $ at separations of integer multiples of the central source's FWHM \citep{furlan2017}. The brightness of each injected source was scaled until standard aperture photometry detected it with $5\sigma $ significance. The resulting brightness of the injected sources relative to the target set the contrast limits at that injection location. The final $5\sigma $ limit at each separation was determined from the average of all of the determined limits at that separation and the uncertainty on the limit was set by the rms dispersion of the azimuthal slices at a given radial distance. The sensitivity curve is shown in Figure \ref{fig:AO} along with a `zoomed' inset image centered on the primary target showing no other companion stars.

\section{Analyses and Results}
\label{sec:analysis}

\subsection{Stellar properties}
\label{sec:host_star}

\subsubsection{Spectroscopic properties and kinematics}
Comparing the spectrum obtained by ALFOSC with the empirical spectral templates of \cite{2017ApJS..230...16K}, we determined the spectral type of the host star to be M0V with an uncertainty of 0.5 subtype. The M0 template with solar metallicity provides a good match (Figure~\ref{TOI-1749-spectrum-not_alfosc}), although there are slight differences in the depths of the TiO bands that may hint at small metallicity deviations ($\Delta$[Fe/H] $\approx \pm 0.5$ dex) from the solar composition. While the ALFOSC spectrum does not allow us to measure the metallicity of the host star with high precision, a very low metallicity of [Fe/H] $<-1.0$ can be ruled out from the fact that the spectrum shows no strong absorption features of hydrides. The spectrum also shows no strong evidence for stellar chromospheric activity (e.g., H$\alpha$ and the red Ca\,{\sc ii} triplet are seen in absorption), indicating that the star is neither very active nor young.

The Galactic space velocities $UVW$ of TOI-1749 were derived using the Gaia coordinates and proper motions listed in Table~\ref{tab:stellar_properties}. We confirmed the Gaia radial velocity, $v_r$ = $-1.8 \pm 2.2$ km\,s$^{-1}$, by determining the star's radial velocity with the ALFOSC spectrum. The centroids of various atomic lines of Ca\,{\sc ii}, Fe\,{\sc i}, K\,{\sc i}, and Ti\,{\sc i} were compared to their laboratory air wavelengths and the mean observed velocity was corrected for the diurnal and lunar velocities and the motion of the Earth--Moon barycenter around the Sun. The ALFOSC heliocentric velocity is $v_r = -4.0 \pm 8.8$ km\,s$^{-1}$. We employed the Gaia radial velocity because of its smaller uncertainty and contrasted velocity zero-points to calculate the $U$, $V$, and $W$ heliocentric velocity components in the directions of the Galactic center, Galactic rotation, and north Galactic pole, respectively, with the formulation developed by \citet{johnson87}. Note that the right-handed system is used and that we did not subtract the solar motion from our calculations. The uncertainties associated with each space velocity component were obtained from the observational quantities and their error bars after the prescription of \citet{johnson87}. The resulting three space velocities are  given in Table~\ref{tab:stellar_properties}. TOI-1749 has kinematics typical of ``the field'' indicating a likely age of $>$ 0.8 Gyr.

\subsubsection{Photometric properties}

\begin{figure}[htp]
    \centering
    \includegraphics[width=8.6cm]{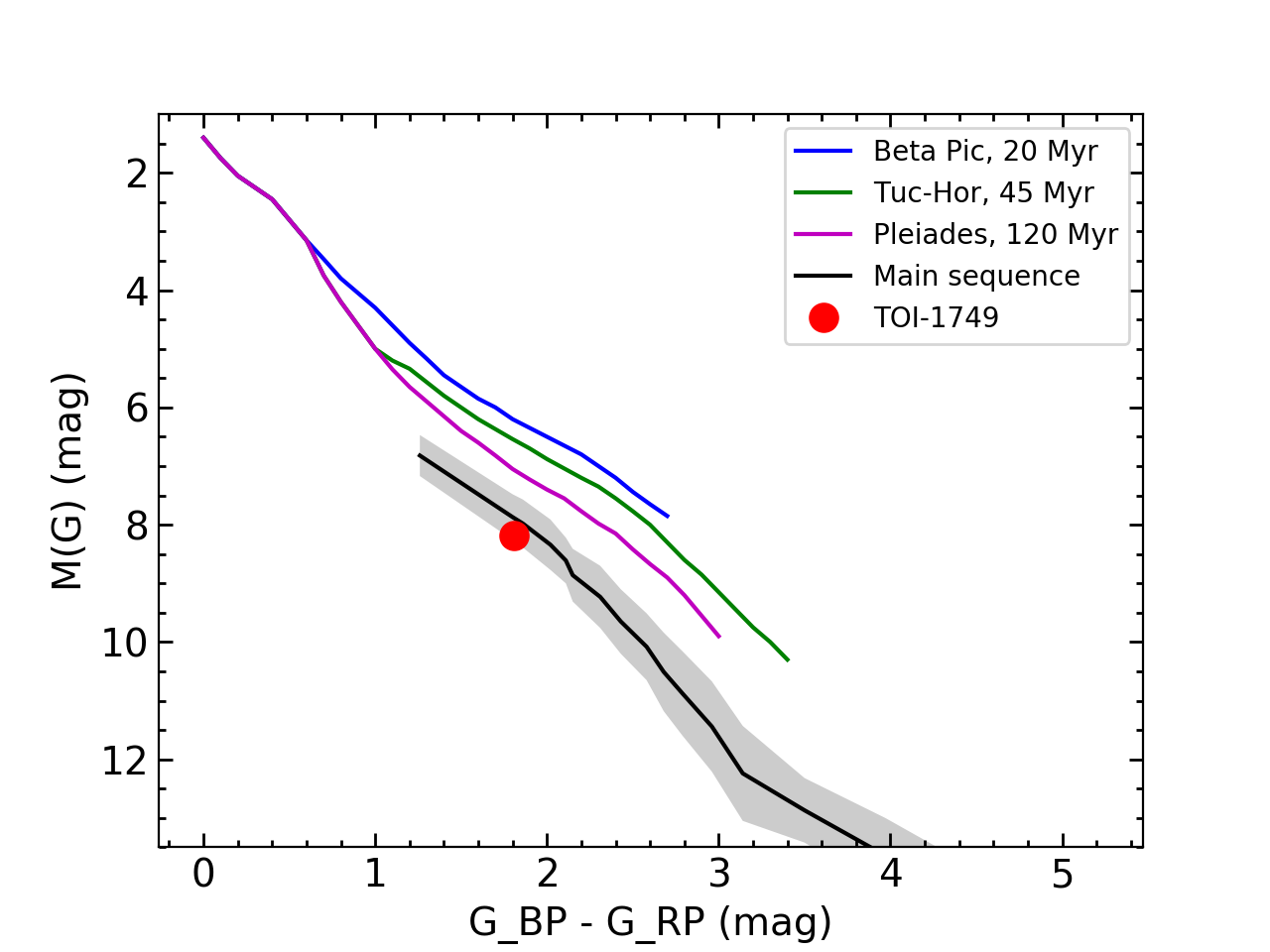}
    \caption{The location of TOI-1749 (red circle) in the Gaia absolute magnitude vs. $G_{BP} - G_{RP}$ color plane, which indicates that it is a main-sequence M0-type star. The young stellar tracks of the $\beta$ Pic (blue) and  Tucana-Horologium (green) moving groups and the Pleiades cluster (purple) are taken from \citet{2018AJ....156..271L} and the main sequence of field stars (black) from \citet{2020A&A...642A.115C}. The gray area represents the dispersion observed among stars of the main sequence.}
    \label{gaia_diagram}
\end{figure}

The spectral type and age of TOI-1749 estimated from the ALFOSC spectrum are also supported by the location of TOI-1749 in the Gaia color-magnitude diagram (CMD) displayed in Figure~\ref{gaia_diagram}; this star has colors and absolute magnitudes compatible with its being a main-sequence M0 star. All stellar sequences shown in Figure~\ref{gaia_diagram} were built using Gaia photometry and parallaxes (see \citealt{2018AJ....156..271L} for the young isochrones and  \citealt{2020A&A...642A.115C} for the main-sequence track). Note that TOI-1749's absolute $G$-band magnitude is slightly fainter than that of the main sequence for its color, which is consistent with a slightly sub-solar metallicity.

\begin{figure}[htp]
    \centering
    \includegraphics[width=8.6cm]{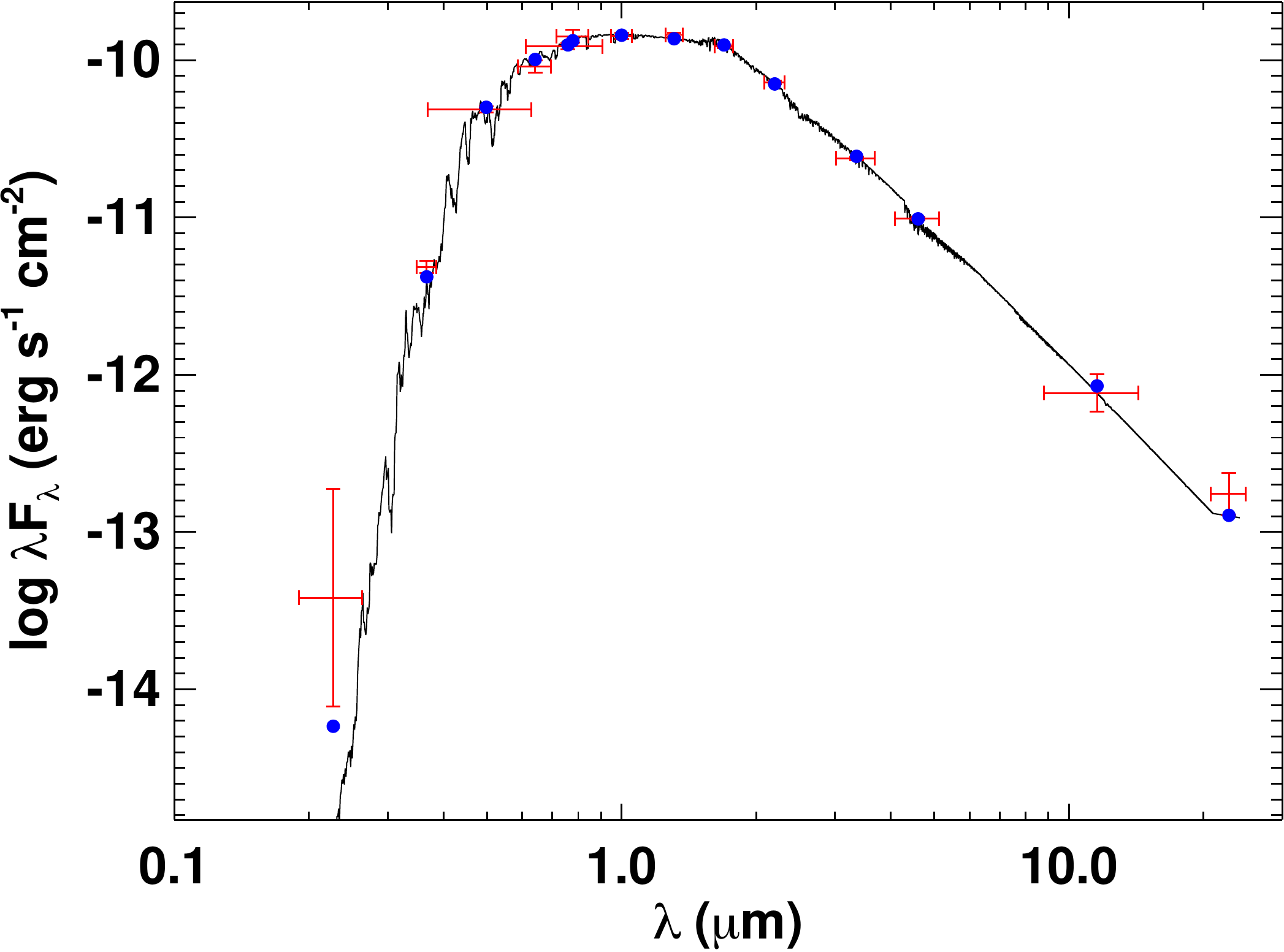}
    \caption{Spectral energy distribution of TOI-1749. Red symbols represent the observed photometric measurements, where the horizontal bars represent the effective widths of the passbands. Blue symbols represent the model fluxes from the best-fit NextGen atmosphere model (black).}
    \label{fig:SED}
\end{figure}

We estimated the physical parameters of the host star empirically via the spectral energy distribution (SED) constructed from  catalog broadband photometric magnitudes, according to the methodology described in \citet{StassunTorres:2016,Stassun:2017}. The available photometry includes the near-UV magnitude from GALEX, the $G G_{\rm BP} G_{\rm RP}$ magnitudes from Gaia, the $ugri$ magnitudes from SDSS and Pan-STARRS, the $JHK_S$ magnitudes from 2MASS, and the W1--4 magnitudes from WISE, as shown in Figure~\ref{fig:SED}. In this fitting, we imposed a Gaussian prior of [Fe/H] = $-0.05 \pm 0.20$~dex on the metallicity, which represents the metallicity distribution of nearby M dwarfs \citep{2014ApJ...791...54G}, and we also fit for $T_{\rm eff}$ and $\log g$. We also assumed no interstellar extinction due to the proximity of the star \rev{(100~pc)}. 
\rev{Note that according to the 3D extinction map of \cite{2019ApJ...887...93G} \footnote{\url{http://argonaut.skymaps.info}}, the reddening along the line of sight toward the minimum reliable distance of 274~pc is estimated to be $E(g-r)=0.009 \pm 0.009$, which justifies the above assumption.}
From this analysis, we obtained $T_{\rm eff} = 3985 \pm 55$~K, $\log g = 4.70 \pm 0.05$, and [Fe/H]=$-0.26 \pm 0.08$~dex. 

According to the TESS Input Catalog ver. 8 \citep[TICv8,][]{Stassun_2019}, the mass and radius of TOI-1749 are $M_s = 0.555 \pm 0.020 M_\odot$ and $R_s = 0.561 \pm 0.017 R_\odot$, respectively, which were calculated using empirical mass-$M_K$ \citep{2019ApJ...871...63M} and radius-$M_K$ \citep{2015ApJ...804...64M} relations for M dwarfs, respectively, where $M_K$ is the $K$-band absolute magnitude. 
Here, we updated these estimates using the revised distance from Gaia DR2 and the [Fe/H] value estimated from the SED fitting, obtaining $M_s = 0.58 \pm 0.03 M_\odot$ and $R_s = 0.55 \pm 0.03 R_\odot$.
The derived physical parameters of TOI-1749 are summarized in Table \ref{tab:stellar_properties}.

\subsubsection{Stellar variability}
\label{sec:rotation}

We searched for periodic variability of TOI-1749 associated with stellar rotation \rev{in the TESS PDC-SAP light curves of all available sectors by a Generalized Lomb-Scargle (GLS) periodogram, but found no significant periodicity. We also investigated}
the $g$- and $r$-band light curves taken from the public data release 5 (DR5) of the Zwicky Transient Facility \citep[ZTF;][]{Masci_2018}, which, after removing bad-quality data points, consist of 572 and 589 data points with photometric \rev{dispersions} of 0.013 mag and 0.010 mag \rev{in rms}, respectively, spanning 32 months between 2018 March and 2020 November. 
Again we found no significant periodic signals in \rev{GLS} periodograms, ruling out the presence of photometric variability with semi-amplitudes of 9~mmag and 6~mmag (at 3 $\sigma$ confidence level) in $g$ and $r$ bands, respectively. The absence of significant stellar variability is consistent with the spectroscopic diagnostic that the star is not young and active. 

\subsection{TESS light curve}

\subsubsection{Searching for a third planetary candidate}
\label{sec:TESS_03}

The candidate transit signals of TOI-1749c and TOI-1749d were first identified from the first six sectors (Sectors 14--19) by the SPOC pipeline, and were later confirmed with additional six sectors (Sectors 20--21 and 23--26). 
However, no additional threshold crossing event (TCE) was detected by the pipeline, not even from all the 12 sectors.
We confirmed the transit signals of TOI-1749c and TOI-1749d in the PDC-SAP light curve using the transit least squares (TLS) method \citep{2019A&A...623A..39H}, yet failed again to find an additional planetary signal in the residual light curve from which the transit signals of TOI-1749c and d were subtracted.

To further search for additional planetary candidates, we removed systematic trends in the residual PDC-SAP light curve that are apparent in several sectors. To model the systematic trends, we applied a Gaussian process (GP) implemented in {\tt celerite} \citep{2017AJ....154..220F} with a stochastically driven, damped simple harmonic oscillator (SHO) as a kernel function. The power spectral density of SHO is written as
\begin{eqnarray}
S(\omega) = \sqrt{\frac{2}{\pi}} \frac{S_0 \omega_0^4}{(\omega^2 - \omega_0^2)^2 + \omega_0^2\omega^2/\mathcal{Q}^2},
\end{eqnarray}
where $\omega_0$ is the frequency of undamped oscillation, $S_0$ is a scale factor to the amplitude of the kernel function, and $\mathcal{Q}$ is a quality factor \citep{1990ApJ...364..699A,2017AJ....154..220F}. This model can capture damping oscillatory behaviors with a characteristic frequency of $\omega_0 \sqrt{1 - 1/4\mathcal{Q}^2}$ if $\mathcal{Q} > 1/2$ \citep{2017AJ....154..220F}. 
Because the amplitudes and timescales of the systematics can be different from sector to sector, we independently modeled $S_0$ and $\omega_0$ for each sector, while fixing $\mathcal{Q}$ at unity for all sectors to avoid over fitting. 

\begin{figure*}[htp]
    \gridline{
        \fig{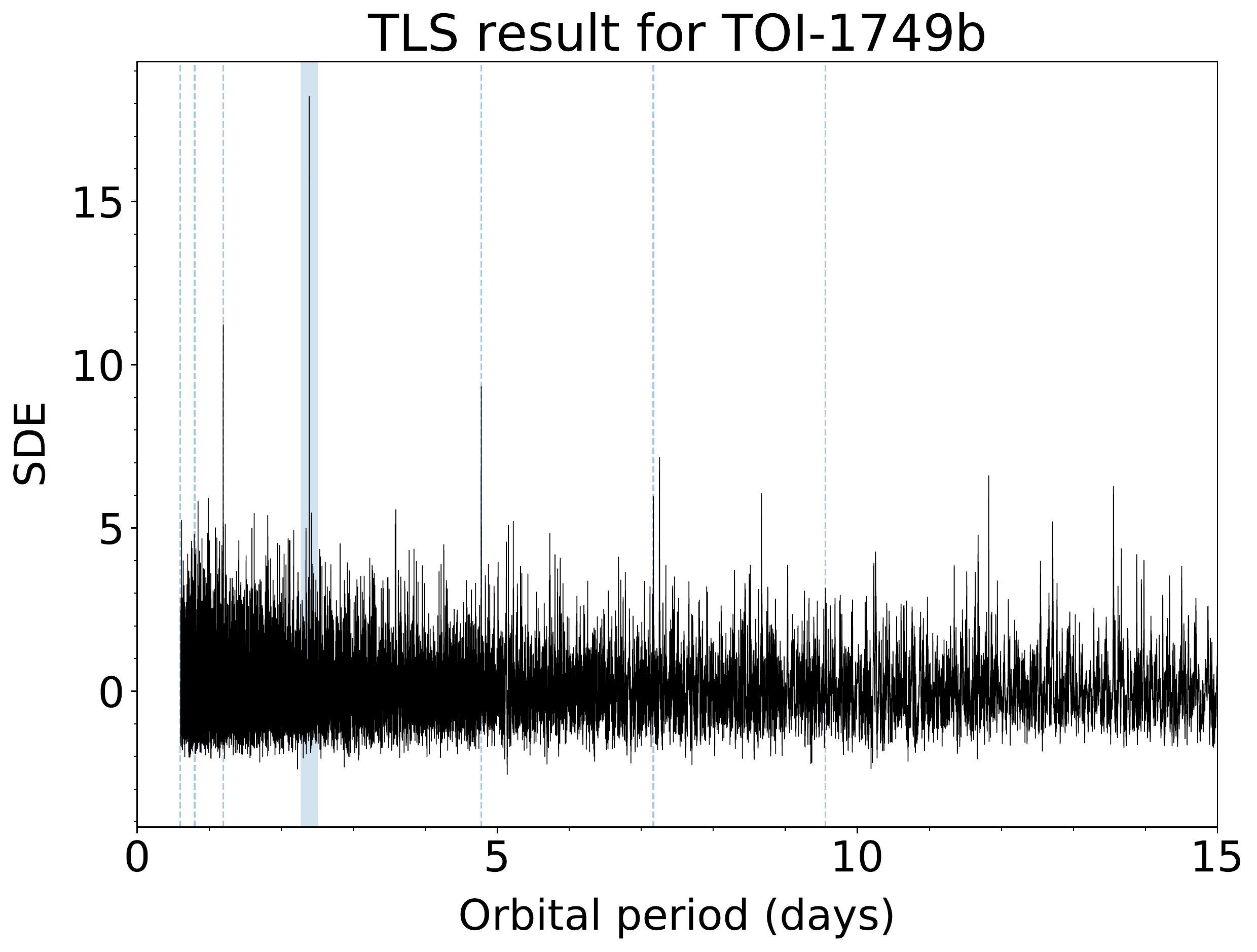}{0.4\textwidth}{}
        \fig{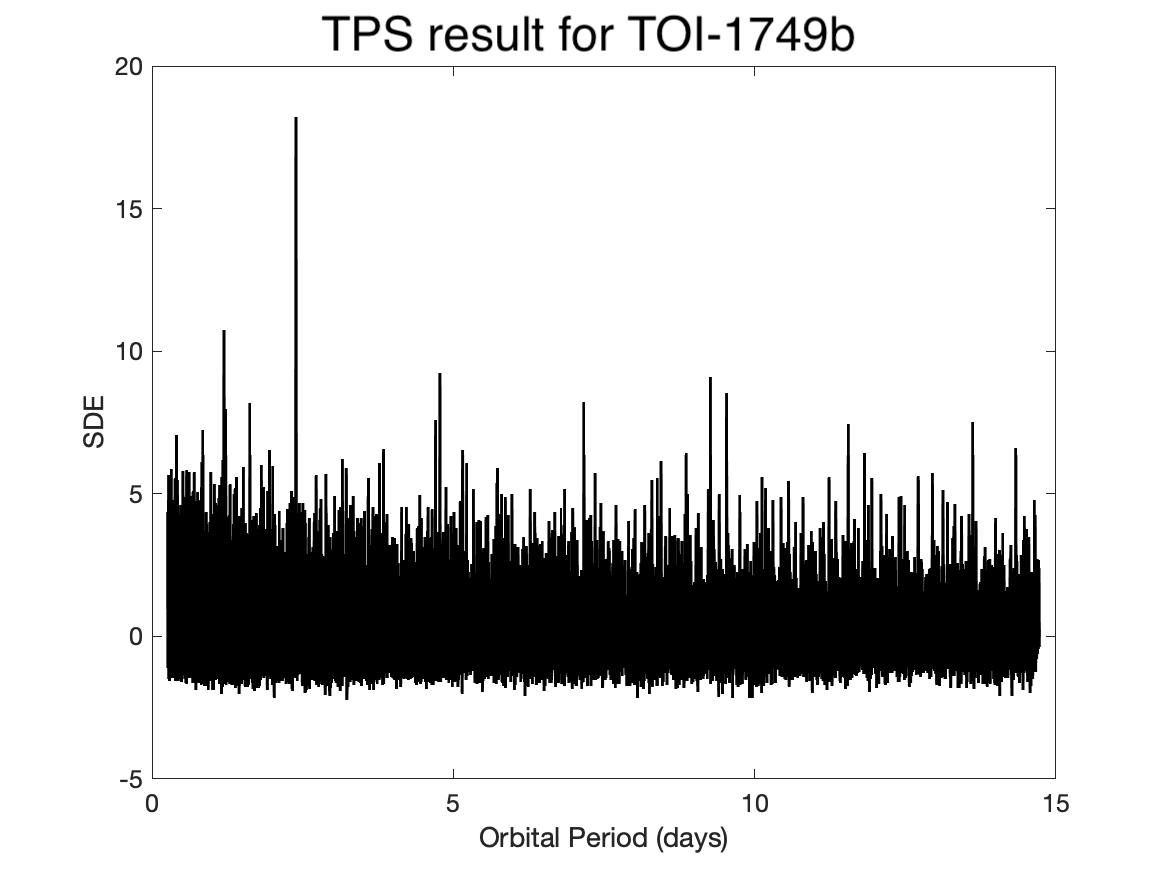}{0.4\textwidth}{}
    }
    \caption{Left: SDE vs. trial orbital period returned by the TLS analysis for TOI-1749b. The SDE reaches a peak of 18.2 at a period of 2.389 days. Right: same as the left panel, but from the TPS analysis, where SDE is calculated as per \cite{2019A&A...623A..39H}.}
    \label{fig:periodogram}
\end{figure*}

After removing the systematic trends using the Gaussian process model with the derived median values, we performed the TLS analysis again, finding an additional transit signal with an orbital period of 2.389 days, \rev{as shown in the left panel of Figure \ref{fig:periodogram}.}
This signal has a signal detection efficiency (SDE) of 18.2, which corresponds to a false alarm probability of $8 \times 10^{-5}$. 

The same transit signal was also detected with an independent pipeline of Open Transit Search ({\tt OpenTS}\footnote{https://github.com/hpparvi/opents}) with a signal-to-noise (S/N) ratio of the transit signal of 4.8.

Motivated by the detection of this marginal signal of a third planet, we revisited the transit signal search by the SPOC\rev{’s Transiting Planet Search (TPS) module}.
This \rev{module} iteratively searches a light curve for transiting planet signatures until no TCE can be returned \citep{2018PASP..130f4502T,2019PASP..131b4506L}, recording the strongest signals even when none meet the detection criteria. Based on all available data in sectors 14--26, we found that the strongest signal for a third transiting planet was consistent with the signal detected by the TLS and OpenTS analyses, with a multiple event detection statistic (MES) of 7.02. This \rev{maximum} MES value was just \rev{over 1\%} below the transit detection threshold for the TESS transit search of 7.1, and was not initially sent to the SPOC Data Validation (DV) process.
Note that we estimate the SDE of the signal detected by TPS as 18.2 (as shown in the right panel of Figure \ref{fig:periodogram}), which is identical to the SDE value of the signal detected by the GP+TLS analysis\footnote{The metric used by TLS is the change in $\chi^2$ for the model with the transiting planet signature at the trial orbital period, depth and duration compared to a no-planet model, and this corresponds to the square of the linear detection statistic used by classical detection algorithms such as TPS.}. This coincidence can be understood as follows. Both TPS and GP+TLS explicitly account for the correlation structure of the observation noise (residual systematic errors plus stellar variability).  GP+TLS explicitly models the correlation structure of the observation noise by construction. That said, TPS models the Power Spectral Density (PSD) of the observation noise process using a nonparametric model via an adaptive, wavelet-based matched filter \citep{2002ApJ...575..493J}. TPS thus performs a joint noise-characterization/signal detection task that explicitly accounts for time-varying, non-white observation noise in the formulated detector. Since the PSD is defined as the Fourier Transform of the correlation function of a wide-sense stationary process \citep{1999ITSP...47...10K}, the approach implemented in TPS can be viewed as the dual of the use of a GP to model the light curve prior to running a search with TLS on the residual light curve.

We then ran the DV module in a supplemental mode for the candidate planets in TOI-1749 including the third planetary candidate, TOI-1749b. Based on a limb-darkened transiting planet model fit \citep{2019PASP..131b4506L} in the DV module, the signal-to-noise (S/N) ratio of the transit signal of TOI-1749b was calculated as 7.5.
Furthermore, the transit signal was fully consistent with that of a transiting planet for all supplemental DV diagnostics: odd/even transit depth consistency test, difference image centroid offset analysis, and optical ghost diagnostic test \citep{2018PASP..130f4502T}.

Given that this candidate signal was detected by three independent analyses and passed the above DV tests, we consider TOI-1749b as a robust planetary candidate and include it in the subsequent analyses.
It should be noted, however, that although the difference image centroid offsets indicated that the location of the transit source was consistent with that of the target star, they were unable to eliminate all nearby stars as the source of the transit signal.
We statistically validate that the target star is the source of the transit signal in Section \ref{sec:validation}. This is also supported by a marginal detection of a transit signal of TOI-1749b on target from the MuSCAT2 observation as described in Section \ref{sec:trfit_GR_03}.

\subsubsection{Transit model fit}
\label{sec:trfit_TESS}

To estimate the transit parameters of the three planetary candidates, we first modeled the PDC-SAP light curve with transit models assuming that all three planetary candidates have constant orbital periods.
We modeled the transit light curves with a Mandel \& Agol model implemented by {\tt PyTransit} \citep{Parviainen2015} with the following parameters: scaled semi-major axis $a/R_s$, impact parameter $b$, planet-to-star radius ratio $R_p/R_s$, eccentricity $e$, longitude of periastron $\varpi$, orbital period $P$, reference transit time $t_0$, and two coefficients $u_1$ and $u_2$ for the stellar limb-darkening effect for which we assume a quadratic limb-darkening law. 
We assumed a circular orbit for all planets by fixing $e$ and $\varpi$ to zero. In addition, we placed informative Gaussian priors on $a/R_s$, $u_1$, and $u_2$ as follows. For the mean value and standard deviation of $a/R_s$, we utilized the stellar density estimated from $M_s$ and $R_s$ derived in Section \ref{sec:host_star}, $\rho_s = 4.91 \pm 0.84$~g~cm$^{-3}$, which was converted to $a/R_s$ for the circular orbits of TOI-1749b, TOI-1749c, and TOI-1749d as $a/R_s = 11.40 \pm 0.65$, $17.35 \pm 0.99$, and $27.7 \pm 1.6$, respectively. For $u_1$ and $u_2$, we adopted $u_1 = 0.320 \pm 0.012$ and $u_2 = 0.255 \pm 0.017$, which were calculated by {\tt LDTk} \citep{Parviainen2015b} based on \rev{PHOENIX} stellar models \citep{Husser2013} for the range of stellar parameters estimated in Section \ref{sec:host_star}. Note that we enlarged the uncertainties of $u_1$ and $u_2$ provided by {\tt LDTk} by a factor of 3 considering the systematic differences between stellar models \rev{and calculation methods for a given set of stellar parameters found in the tabulated values of \cite{2017A&A...600A..30C}.} We assumed uniform priors for the other parameters. 
Simultaneously with the transit models, we also modeled systematic trends in the PDC-SAP light curve using the Gaussian process model in the same way as in Section \ref{sec:TESS_03}.

To estimate the posterior distributions of the total of 53 free parameters, we ran a Markov Chain Monte Carlo (MCMC) analysis using {\tt emcee} \citep{2013PASP..125..306F}. From an MCMC run with 106 walkers and $2\times10^4$ steps after convergence, we derived the median values and 1$\sigma$ uncertainties of the parameters as listed in Table \ref{tbl:trfit}. 

\begin{figure*}[htp!]
    \centering
    \includegraphics[width=16cm]{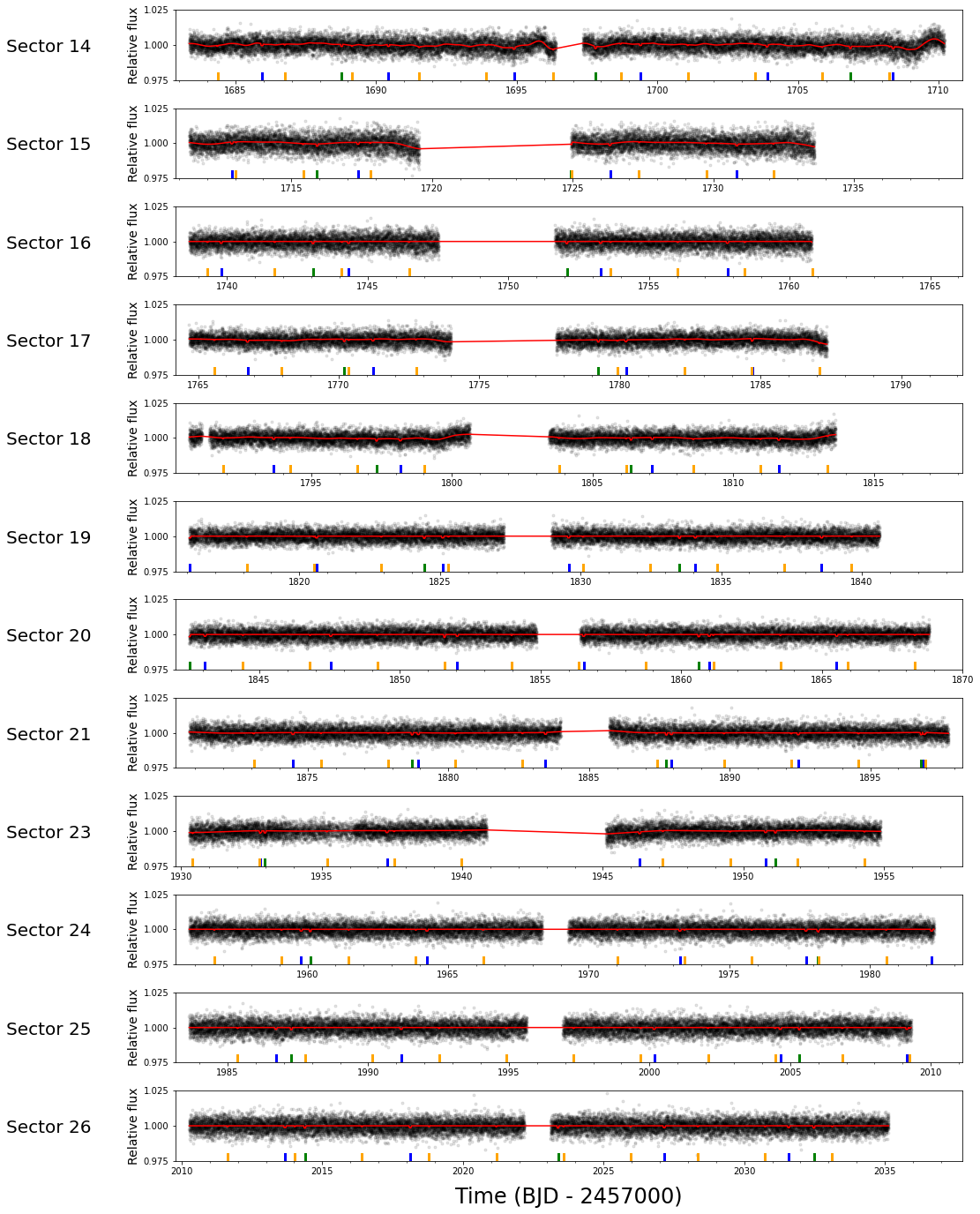}
    \caption{The TESS PDC-SAP light curve of TOI-1749 (black dots) and the best-fit systematics+transit model (red). The timings of individual transits of TOI-1749b, TOI-1749c, and TOI-1749d are marked at the bottom of each panel in orange, blue, and green, respectively.}
    \label{fig:lc_TESS_undetrend}
\end{figure*}

\begin{figure}[htp!]
    \centering
    \includegraphics[width=8cm]{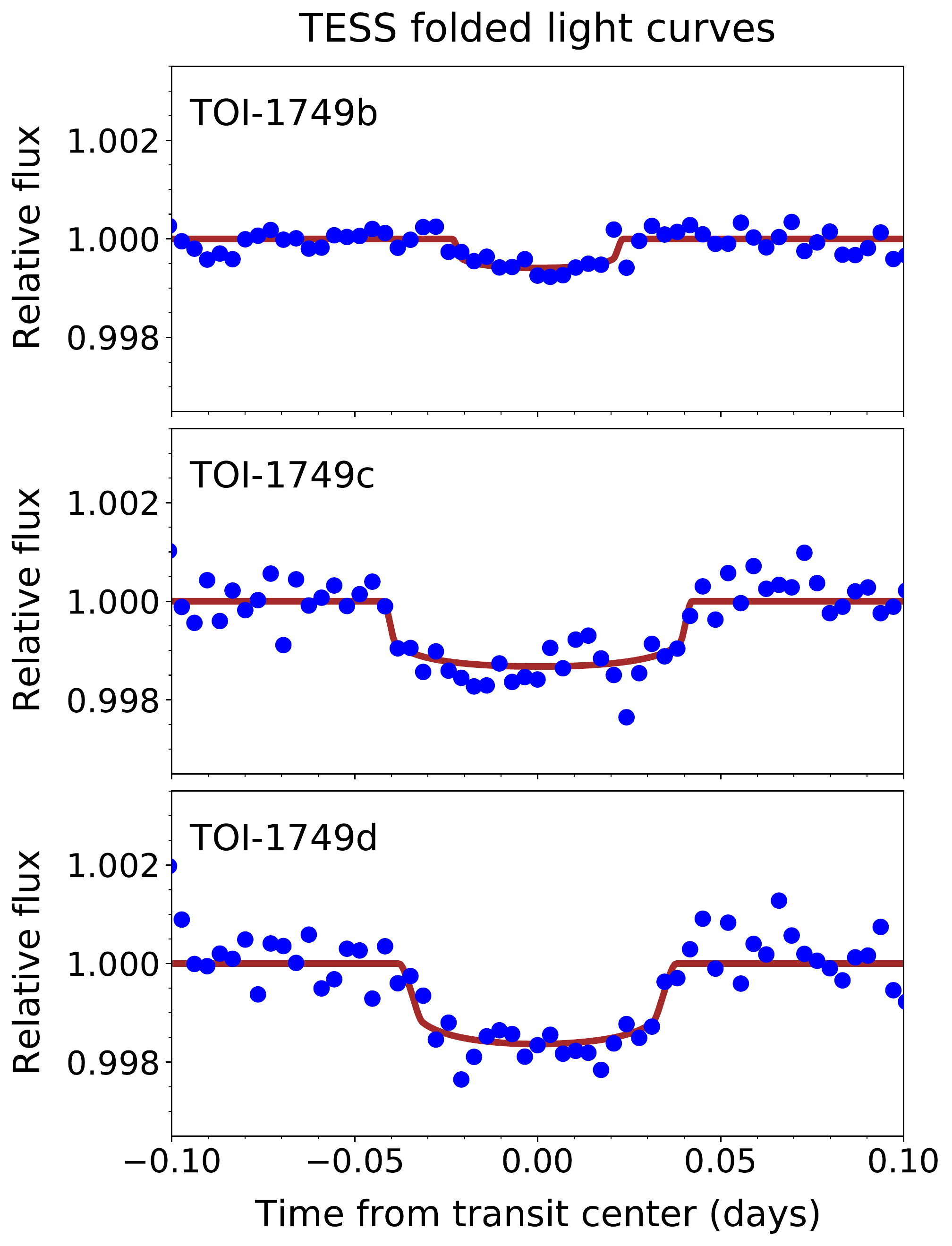}
    \caption{Detrended, phase-folded, and 5 minutes binned PDC-SAP light curves of TOI-1749b, TOI-1749c, and TOI-1749d (blue points). The brown solid lines are the best-fit transit models.}
    \label{fig:lc_TESS_folded}
\end{figure}

Next, to investigate the effect of possible TTVs, we fit the same light curve with the same transit+systematic models, but set aside the assumption of linear ephemerides and, instead, let individual mid-transit times ($T_c$) be free for TOI-1749c and TOI-1749d. Note that because the S/N ratios of individual transits of TOI-1749b are too low to measure $T_c$, we held the assumption of a linear ephemeris for this planetary candidate. Given the number of transits that the TESS data cover of 58 and 28 for TOI-1749c and TOI-1749d, respectively, the number of free parameters in this model is 135. We ran an MCMC analysis 
in the same way as before, but this time with 270 walkers and $4 \times 10^4$ steps. The measured individual transit times are listed in Table \ref{tbl:transit_times}, and the other transit parameters are reported in Table \ref{tbl:trfit}. The best-fit light curve model is shown in red in Figure \ref{fig:lc_TESS_undetrend}, and phase-folded light curves are shown in Figure \ref{fig:lc_TESS_folded}.
The individual-$T_c$ model improves the best $\chi^2$ value by 438.0 with 82 additional free parameters over the linear ephemeris model, for a total of 188,918 data points. 
\rev{Among these data points,}
most of the $\chi^2$ improvement comes from only the data around individual transits, which corresponds to $\sim20,000$ data points. Therefore, the difference of Bayesian information criterion \citep[BIC $\equiv \chi^2 + k \ln N$, where $k$ is the number of free parameters and $N$ is the number of data points;][]{1978AnSta...6..461S} between the two models is estimated to be $\sim-370$. This indicates that the individual-$T_c$ model does not better describe the data over the linear ephemeris model; i.e., we do not detect significant TTV signals in the TESS data from this analysis.
\rev{Note that the same conclusion was obtained by fitting a straight line to the set of derived $T_c$ values and transit epochs, where the uncertainty of $T_c$ was approximated to be a Gaussian; the fit gives a reduced $\chi^2$ of 0.69 and 1.3 for TOI-1749c and d, respectively.}

\begin{deluxetable*}{lcccc}[htp!]
\tablecaption{Results of transit model fitting.}
\label{tbl:trfit}
\tablehead{
\colhead{Parameter} & \colhead{Unit} & \colhead{TESS} & \colhead{TESS} & \colhead{Ground-based}\\
& & (linear ephemeris fit) & (individual $T_{\rm c}$ fit) &\\
}
\startdata
\multicolumn{5}{c}{TOI-1749b}\\
$P$ & days & $2.388821\ ^{+0.000057}_{-0.000073}$ & $2.388843\ ^{+0.000044}_{-0.000076}$ & 2.388843 (fixed)\\
$t_0$ & BJD$_{\rm TDB}$ - 2458680 & $4.3613\ ^{+0.0070}_{-0.0049}$ &  $4.3597\ ^{+0.0066}_{-0.0040}$ &---\\
$a/R_s$ & & $11.53\ ^{+0.61}_{-0.65}$ & $11.60 \pm 0.55$ & 11.60 (fixed) \\
$b$ &  & $0.76\ ^{+0.08}_{-0.15}$ & $0.74\ ^{+0.10}_{-0.16}$ & 0.74 (fixed) \\
$R_p/R_s$ (TESS) &  & $0.0239 \pm 0.0025$ & $0.0241\ ^{+0.0025}_{-0.0022}$ &---\\
$R_p/R_s$ (ground) & & --- & --- & $0.0244\ ^{+0.0056}_{-0.0071}$\\
\hline
\multicolumn{5}{c}{TOI-1749c}\\
$P$ & days & $4.489010\ ^{+0.000040}_{-0.000061}$ & 4.489010 (fixed) & 4.489010 (fixed)\\
$t_0$ & BJD$_{\rm TDB}$ - 2458680 & $5.9386\ ^{+0.0028}_{-0.0019}$ & --- & ---\\
$a/R_s$ & & $17.7 \pm 0.8$ & $17.2 \pm 0.8$ & $17.3 \pm 0.6$\\
$b$ &  & $0.29\ ^{+0.16}_{-0.18}$ & $0.21\ ^{+0.15}_{-0.13}$ & $0.22 \pm 0.12$\\
$R_p/R_s$ (TESS) &  & $0.0344 \pm 0.0011$ & $0.0337 \pm 0.0012$ & ---\\
$R_p/R_s$ ($g$) & & --- & --- & $0.0336\ ^{+0.0039}_{-0.0045}$\\
$R_p/R_s$ ($r$) & &--- & --- & $0.0331 \pm 0.0020$\\
$R_p/R_s$ ($i$) & & --- & --- & $0.0385 \pm 0.0018$\\
$R_p/R_s$ ($z_s$) & & --- & --- & $0.0336 \pm 0.0019$\\
\hline
\multicolumn{5}{c}{TOI-1749d}\\
$P$ & days & $9.04455 \pm 0.00012$ & 9.04455 (fixed) & 9.04455 (fixed)\\
$t_0$ & BJD$_{\rm TDB}$ - 2458680 & $8.7805 \pm 0.0028$ & --- & ---\\
$a/R_s$ & & $28.0 \pm  1.5$ & $28.2 \pm 1.3$ & $27.6 \pm 1.2$\\
$b$ &  & $0.719\ ^{+0.042}_{-0.049}$ & $0.733\ ^{+0.049}_{-0.058}$ & $0.718\ ^{+0.031}_{-0.034}$\\
$R_p/R_s$ (TESS)&  & $0.0387\ ^{+0.0016}_{-0.0018}$ & $0.0399\ ^{+0.0017}_{-0.0022}$ & ---\\
$R_p/R_s$ ($g$) & & --- & --- & $0.0416\ ^{+0.0030}_{-0.0038}$\\
$R_p/R_s$ ($r$) & & --- & --- & $0.0403\ ^{+0.0030}_{-0.0032}$\\
$R_p/R_s$ ($i$) & & --- & --- & $0.0435\ ^{+0.0019}_{-0.0022}$\\
$R_p/R_s$ ($z_s$) & & --- & --- & $0.0447\ ^{+0.0030}_{-0.0032}$\\
\enddata
\end{deluxetable*}

\begin{deluxetable}{lccccc}[htp]
\tablecaption{Mid-times of individual transits.}
\label{tbl:transit_times}
\tablehead{
\colhead{Epoch} & \colhead{$T_c$} & \colhead{1$\sigma$ error} & \colhead{1$\sigma$ error} & \colhead{Instrument} & \colhead{Planet} \\
& (BJD$_{\rm TDB}$)& (upper) & (lower) & &
}
\startdata
0 & 2458685.9405 & 0.0398 & 0.0128 & TESS & c \\
1 & 2458690.4394 & 0.0359 & 0.0735 & TESS & c \\
2 & 2458694.9077 & 0.0089 & 0.0090 & TESS & c \\
3 & 2458699.4013 & 0.0112 & 0.0117 & TESS & c \\
4 & 2458703.9008 & 0.0087 & 0.0263 & TESS & c \\
... &&&&&
\enddata
\tablecomments{Table \ref{tbl:transit_times} is published in its entirety in the machine-readable format. A portion is shown here for guidance regarding its form and content.}
\end{deluxetable}

\subsection{Ground-based light curves}

\subsubsection{TOI-1749c and TOI-1749d}
\label{sec:trfit_GR_01_02}

We modeled the light curves of TOI-1749c and TOI-1749d obtained from the ground-based follow-up observations as follows.
All of the photometric data sets of each planetary candidate that show significant transit signals (check-marked in Table \ref{tbl:observing_log}) were simultaneously fit with transit+systematic models. For the transit model, we used the same one described in Section \ref{sec:trfit_TESS}, but allowed $R_p/R_s$ to be free for each band to check for a chromaticity dependence in this parameter, which can be a sign of flux contamination in the photometric aperture. We let $T_c$ be free for each transit epoch, while treating $b$ and $a/R_s$ as common parameters for all light curves. The limb-darkening parameters of $u_1$ and $u_2$ were fixed for each band to the theoretical values given by {\tt LDTk}, specifically, $(u_1, u_2) = (0.656, 0.129)$, $(0.541, 0.183)$, $(0.374, 0.248)$, and $(0.302, 0.253)$ for the $g$, $r$, $i$, and $z_s$ bands, respectively. 

We modeled the systematics in each light curve by a combination of a linear function of $\Delta X$ and $\Delta Y$, which are the stellar displacements on the detector in the $X$ and $Y$ directions, respectively, and a Gaussian process model \rev{as a function of time} with an approximated Mat\'{e}rn 3/2 kernel implemented in {\tt celerite}. The former function takes account of systematics originating from stellar movements on the detector, which are typically within a few pixels during each transit observation.
When stars move on the detector, they are subject to photometric systematics caused by imperfections in the flat-field corrections. 
The Gaussian process models other nonoscillating, time-correlated noise, which presumably mostly originates in effects of observing through the Earth's atmosphere. The approximated Mat\'{e}rn 3/2 kernel is written by 
\begin{eqnarray}
k(\tau) &=& \sigma^2 [ (1+\frac{1}{\epsilon}) e^{-(1-\epsilon)\sqrt{3}\tau/\rho} \nonumber\\
&& \times (1-\frac{1}{\epsilon}) e^{-(1+\epsilon)\sqrt{3}\tau/\rho}],
\end{eqnarray}
where $\tau$ is the distance of two data points in time, and $\sigma$, $\rho$, and $\epsilon$ are coefficients. We fixed $\epsilon$, which is a quality factor for approximation, to a default value of the code of 0.01. Each light curve was modeled by the GP model with a mean function of 
\begin{eqnarray}
\mu = \mathcal{F}_{\rm transit} \times (c_0 + c_x \Delta X + c_y \Delta Y),
\end{eqnarray}
where $\mathcal{F}_{\rm transit}$ is the transit model and $c_0$, $c_1$, and $c_2$ are the coefficients for the linear systematic model. Among the parameters for the systematic model, we forced $\rho$ to be a common parameter for each transit assuming that the timescale of the time-correlated noise in each night is shared among the different bands, while letting $\sigma$, $c_x$, and $c_y$ be free for individual light curves. Note that $c_0$ was obtained for a given set of parameters by taking a median value of $F  / \mathcal{F}_{\rm transit} - (c_x \Delta X + c_y \Delta Y)$, where $F$ is the observed flux. We also modeled a white jitter noise for each light curve, $\sigma_{\rm jitter}$, in the form of $\sigma_{\rm flux} = \sqrt{ \sigma_{\rm calc}^2 + \sigma_{\rm jitter}^2 }$, where $\sigma_{\rm flux}$ is the uncertainty of the flux of each data point and $\sigma_{\rm calc}$ is the theoretical uncertainty calculated by the photometric pipeline.

\begin{figure*}[htp]
    \plotone{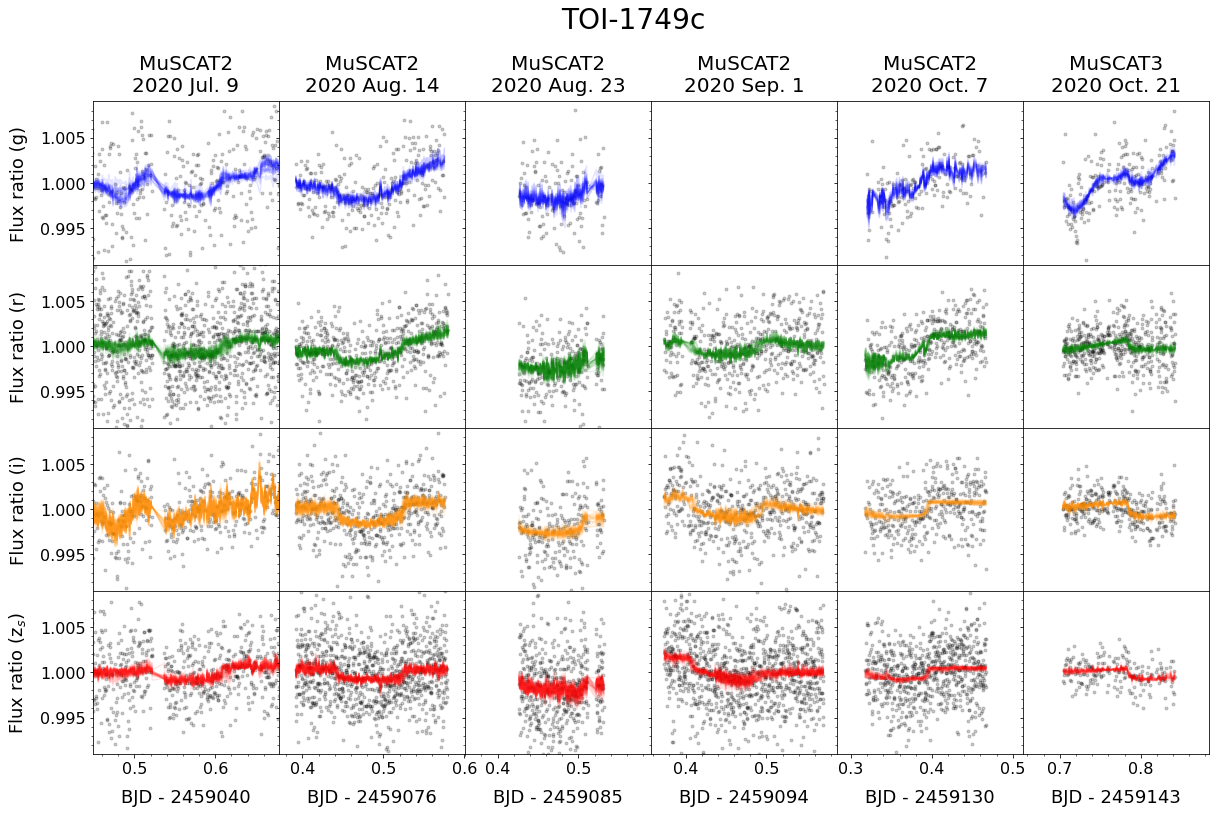}
    \caption{Individual transit light curves of TOI-1749c obtained with ground-based follow-up observations. Black dots and colored lines indicate individual exposure data and 20 randomly selected posterior transit+systematic models (see Section \ref{sec:trfit_GR_01_02}), respectively. Each column shows light curves for a particular night (transit) taken with the instrument indicated at the top of the figure, and the rows show the $g$-, $r$-, $i$-, and $z_s$-band light curves from top to bottom, respectively.}
    \label{fig:lc_GR_TOI-1749.01}
\end{figure*}

\begin{figure*}[htp]
    \plotone{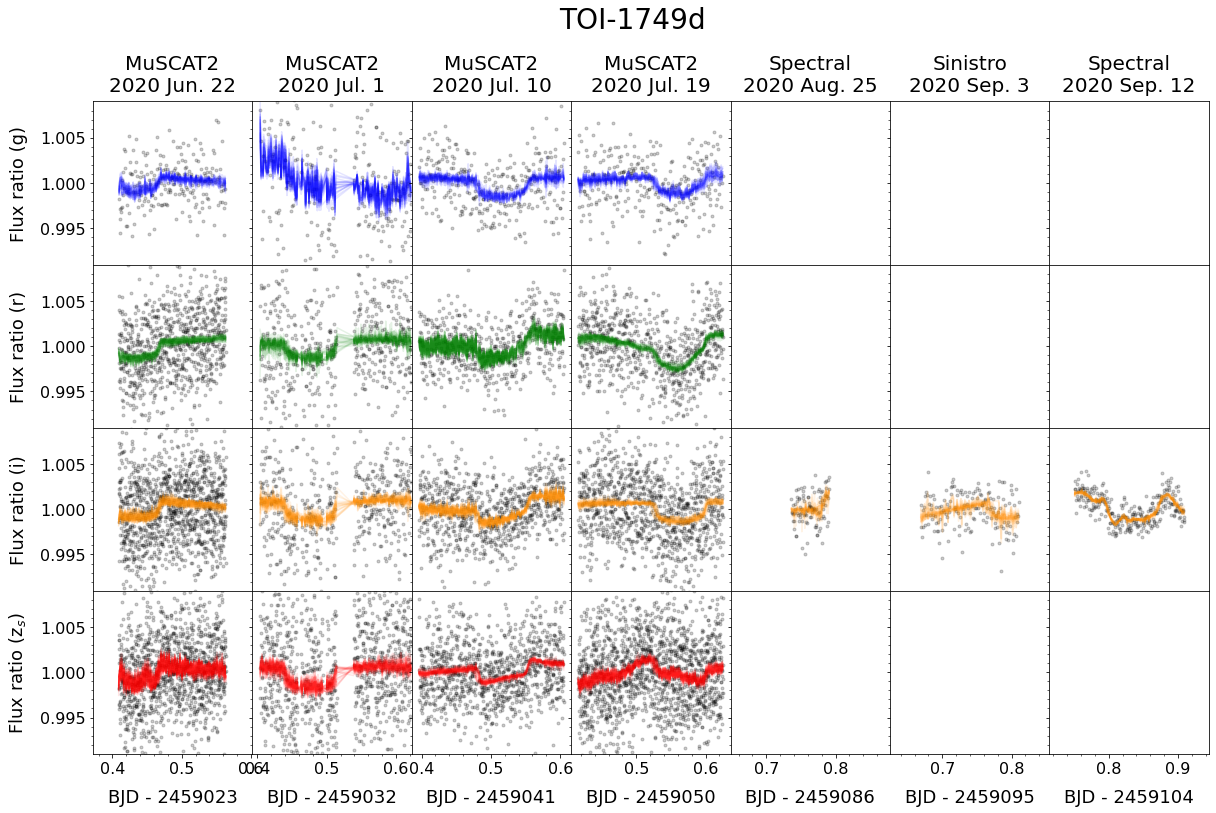}
    \caption{Same as Figure \ref{fig:lc_GR_TOI-1749.01}, but for TOI-1749d.}
    \label{fig:lc_GR_TOI-1749.02}
\end{figure*}

\begin{figure*}[htp]
    \gridline{
        \fig{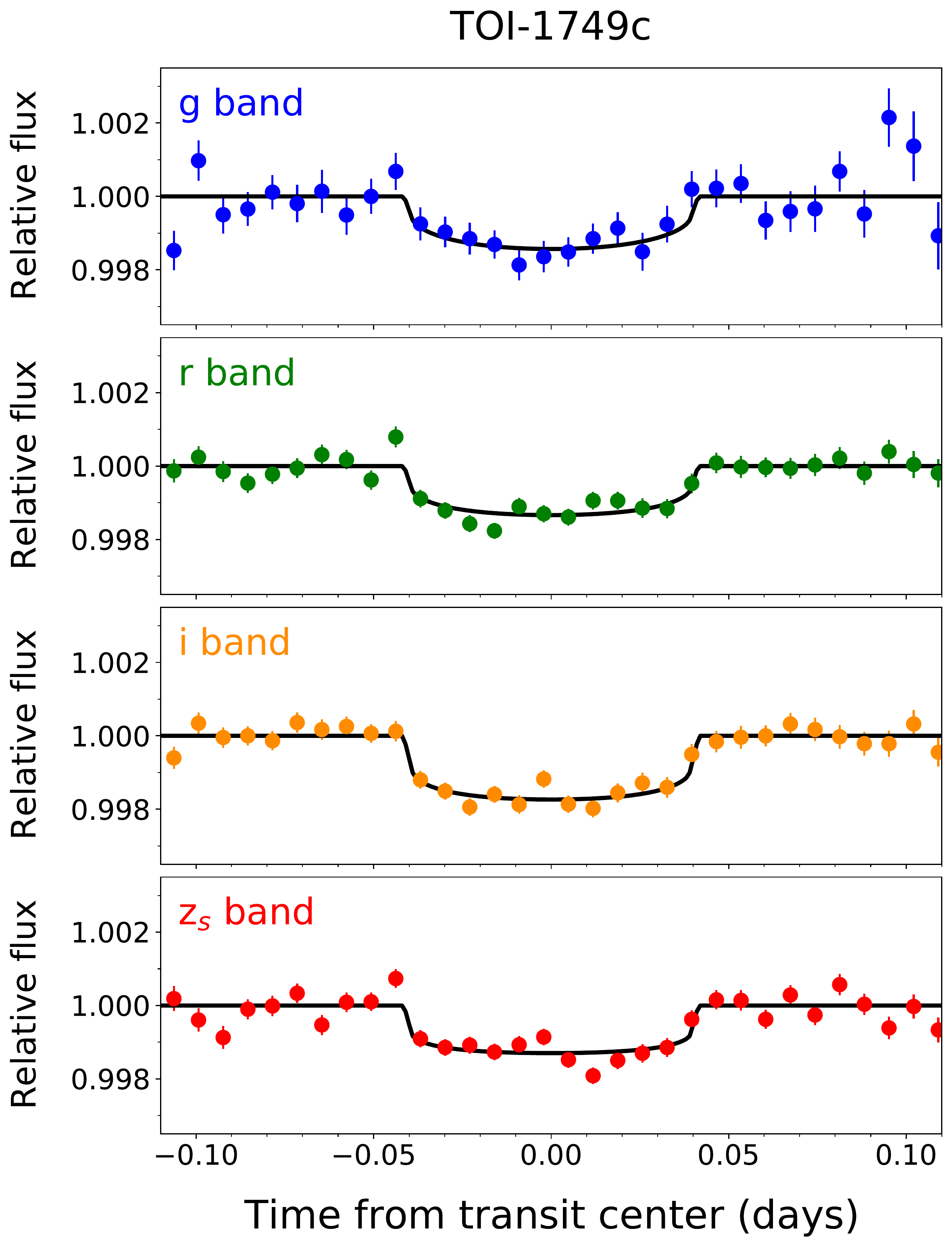}{0.4\textwidth}{}
        \fig{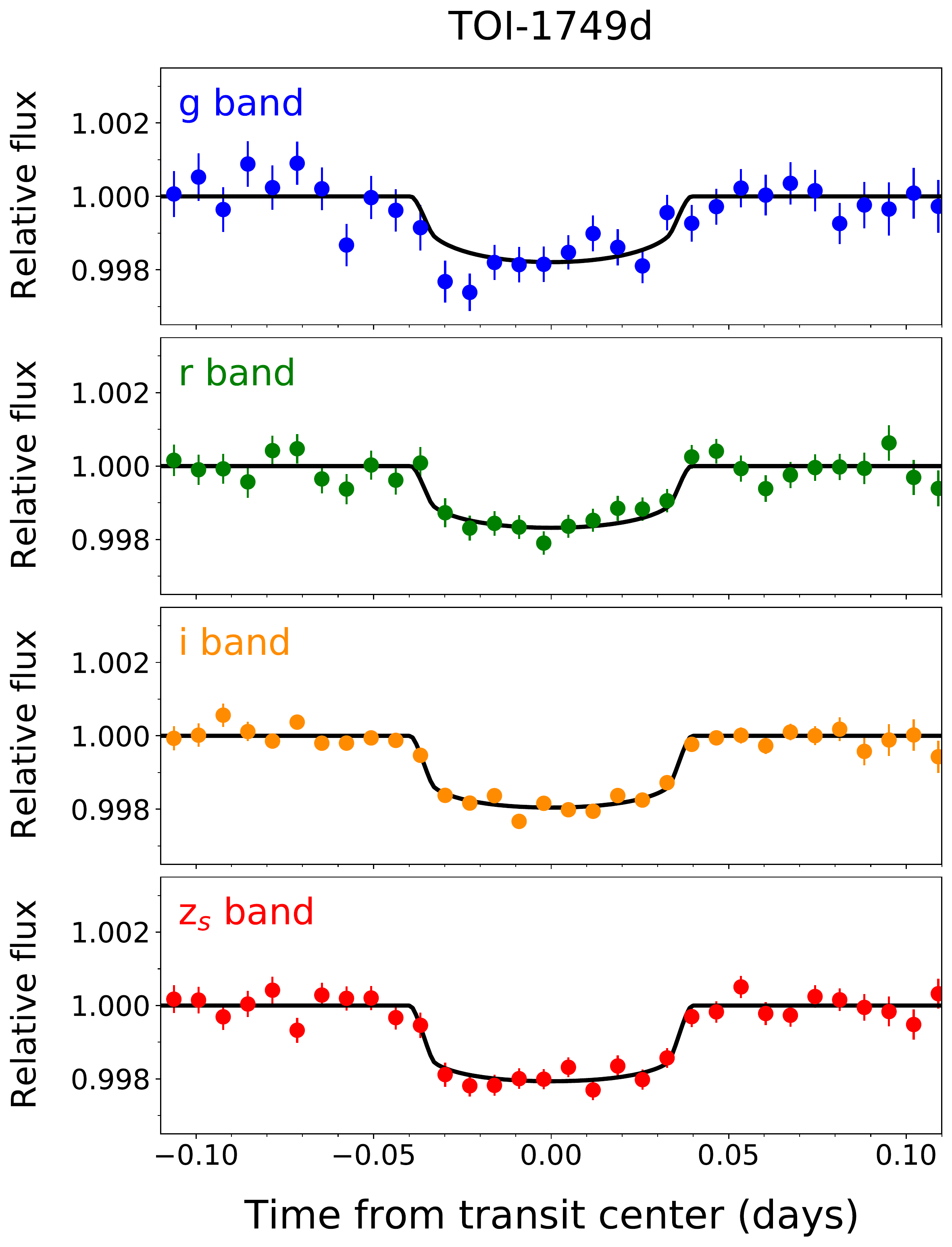}{0.4\textwidth}{}
    }
    \caption{Phase-folded light curves of TOI-1749c (left) and TOI-1749d (right) from the ground-based follow-up observations. Colored circles and black solid lines indicate detrended, 10 minutes binned data and best-fit transit models, respectively.}
    \label{fig:lc_GR_folded}
\end{figure*}

\begin{figure}[htp]
    \centering
    \includegraphics[width=8cm]{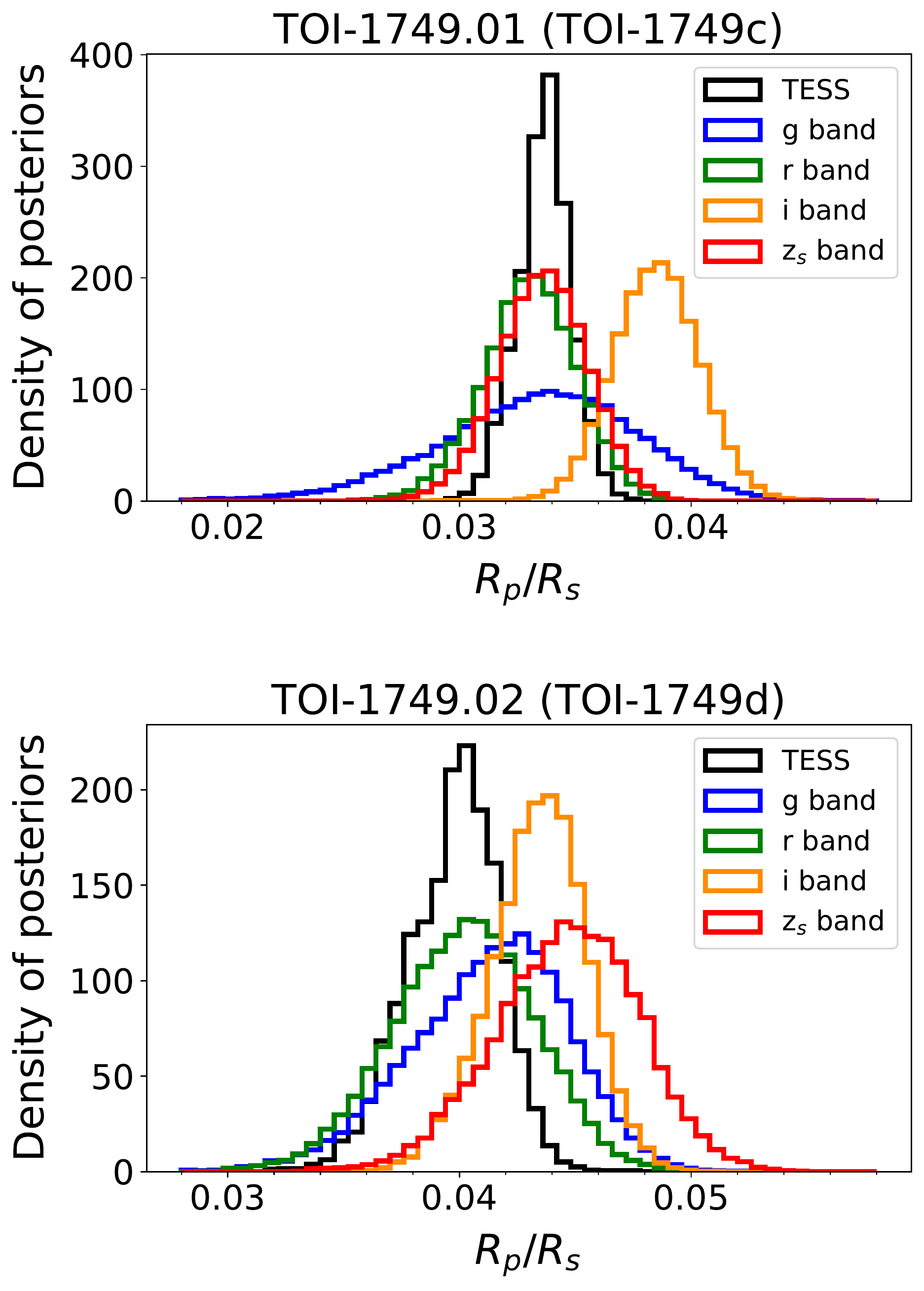}
    \caption{Top: posterior distributions of $R_p/R_s$ for TOI-1749c obtained from the analyses of the TESS light curve (black) and of the ground-based ones (blue, green, orange, and red are for $g$, $r$, $i$, and $z_s$ bands, respectively). Bottom: same as the top panel, but for TOI-1749d. }
    \label{fig:hist_k}
\end{figure}

With this model, we ran MCMC for a total of 23 light curves (6 transit epochs) and 19 light curves (7 transit epochs) for TOI-1749c and TOI-1749d, with numbers of free parameters of 110 and 96, respectively. For $a/R_s$ and $b$, we imposed informative Gaussian priors using the results of the MCMC analysis for the TESS light curve with the individual $T_c$ model; that is, $(a/R_s, b) = (17.2 \pm 0.8, 0.21 \pm 0.14)$ and $(28.2 \pm 0.13, 0.733 \pm 0.053)$ for TOI-1749c and TOI-1749d, respectively. We applied uniform priors for the other parameters, with natural logarithmic form for $\rho$, $\sigma$, and $\sigma_{\rm jitter}$. Using {\tt emcee}, we calculated posterior probability distributions of the parameters from $4 \times 10^4$ MCMC steps with 220 and 192 walkers for TOI-1749c and TOI-1749d, respectively. 
The derived median and 1$\sigma$ confidence intervals of the mid-transit times and the other transit parameters are appended to Tables \ref{tbl:transit_times} and \ref{tbl:trfit}, respectively. 
We show 100 light-curve models randomly selected from the posterior distributions along with the individual ground-based light curves in Figures \ref{fig:lc_GR_TOI-1749.01} and \ref{fig:lc_GR_TOI-1749.02}, and also show the best-fit transit model along with phase-folded, systematic-corrected light curves for each planet and each band in Figure \ref{fig:lc_GR_folded}.
We find that the $R_p/R_s$ values of respective planets measured in five different bands (one from TESS and four from the ground) are consistent with each other within $\sim$2~$\sigma$ as shown in Figure \ref{fig:hist_k}, without any particular trend as a function of wavelength.  We thus find no evidence of flux contamination within the photometric aperture of ground-based observations (2\farcs6--5\farcs2). 
\rev{We note that the atmospheric scale heights of TOI-1749c and TOI-1749d are expected to be $\sim5 \times 10^{-4}$ and $\sim4 \times 10^{-4}$ in units of $R_s$, assuming hydrogen-rich atmospheres with planetary masses of 5 and 7 $M_\oplus$, respectively \citep[the masses predicted using an empirical relation of][]{2017ApJ...834...17C}. Thus, the effect of atmospheric opacity on the measured $R_p/R_s$, which can be $\sim5$ times the scale height, is comparable to or smaller than the measurement uncertainties, unless the planetary masses are unusually small.}

\subsubsection{TOI-1749b}
\label{sec:trfit_GR_03}

\begin{figure*}[htp]
    \gridline{\fig{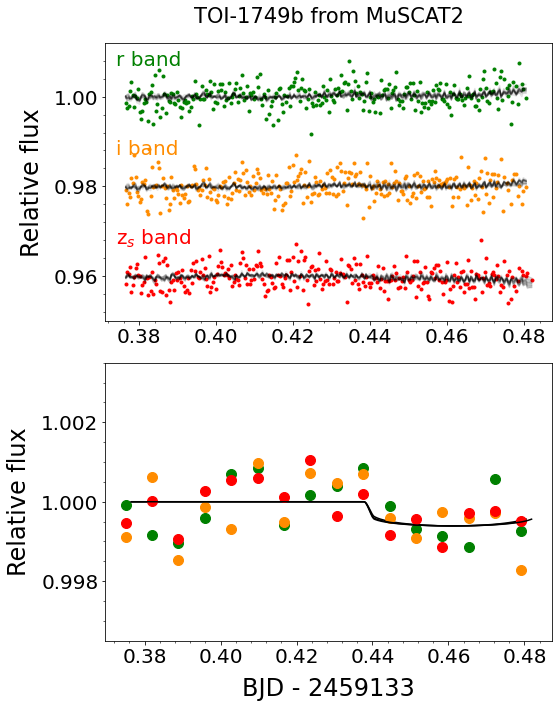}{0.4\textwidth}{(a)}
            \fig{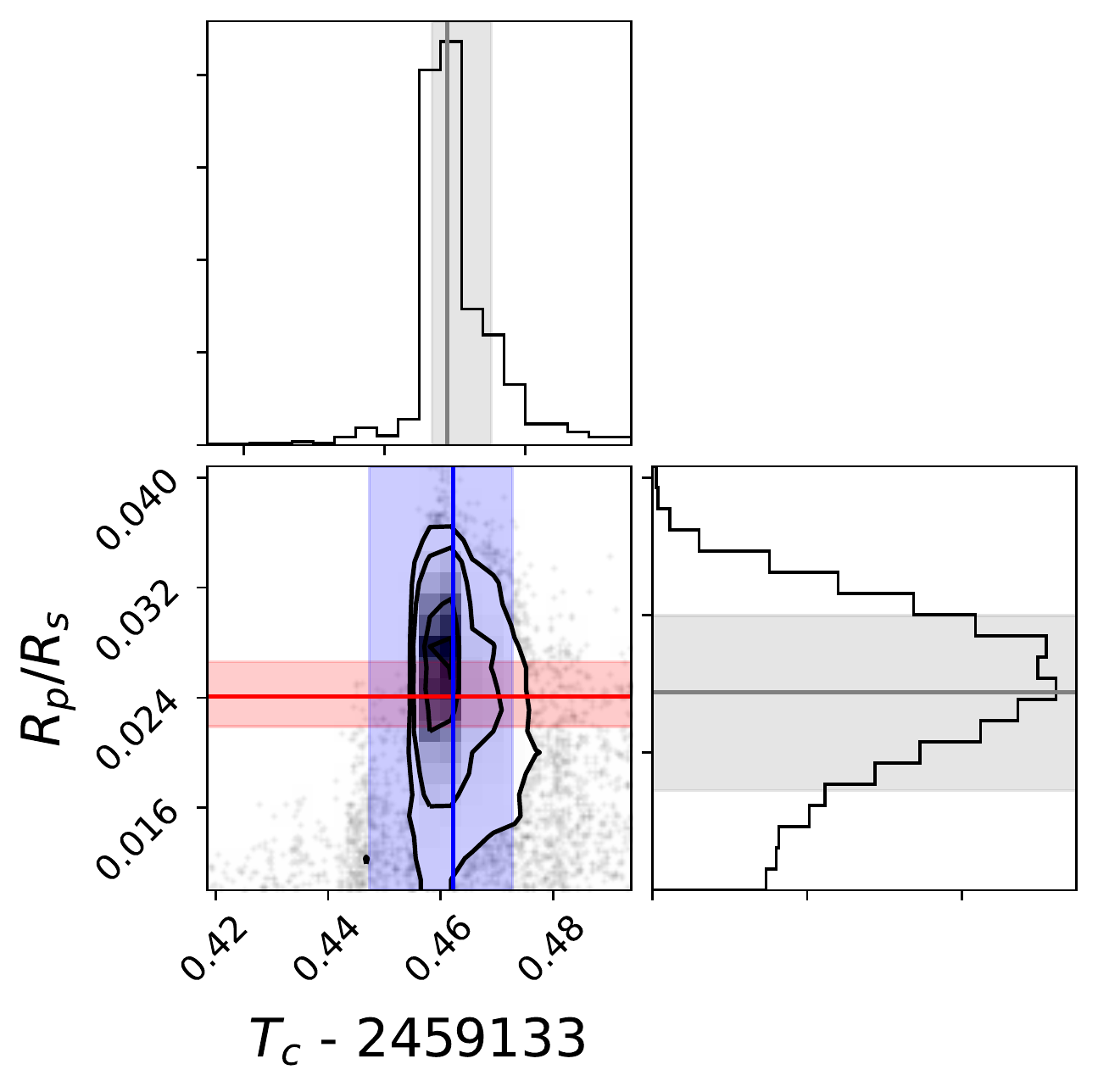}{0.5\textwidth}{(b)}}
    \caption{(a) Top: un-detrended, unbinned light curves of TOI-1749b obtained with MuSCAT2. The $r$-, $i$-, and $z_s$-band data are displayed in green, orange, and red, respectively, with vertical offsets for clarity. Randomly-selected 20 posterior light-curve (transit+systematics) models are shown by black solid lines. Bottom: same as the top panel, but the median systematic model for each band is subtracted from each light curve, the data points in each light curve are binned into 10 minute bins, and the three light curves are overlaid on each other. (b) Bottom left: two-dimensional posterior distributions for $T_c$ and $R_p/R_s$ obtained from the MCMC analysis of the data shown in Panel (a). The red and blue solid lines indicate the median predicted values of $R_p/R_s$ and $T_c$, respectively, from the TESS light curves analyzed in Section \ref{sec:trfit_TESS}. Shaded regions indicate 1$\sigma$ confidence intervals. Top left: histogram of the posterior distribution of $T_c$. The median value and 1~$\sigma$ confidence interval are indicated by a gray solid line and shaded area, respectively. Bottom right: same as the top left panel, but for $R_p/R_s$.}
    \label{fig:TOI-1749.03}
\end{figure*}

The light curves of TOI-1749b obtained with MuSCAT2 show no apparent transit signal, as shown in the top panel of Figure \ref{fig:TOI-1749.03} (a). To see if there is the expected transit signal of this candidate, which has a depth of $\sim$0.058~\% ($\sim$580~ppm) and an expected mid-transit time of $T_c = 2459133.462^{+0.011}_{-0.015}$, we fit a transit+systematics model to the $r$-, $i$-, and $z_s$-band light curves in the same way as in Section \ref{sec:trfit_GR_01_02}, but fixing $b$ and $a/R_s$ to the values obtained from the TESS light curves, i.e., 0.74 and 11.60, respectively. The parameters of $R_p/R_s$ and $T_c$ were shared between the three bands. To avoid sampling unrealistic parameter space, we limited the allowed range of $T_c$ to $\pm 3\sigma$ from the above expected value. We note that the expected TTV amplitude for the transits of this candidate is at most a few minutes, as shown in Section \ref{sec:pdfit}, which is negligible compared to the timing uncertainty from the linear ephemeris. As a result, we find a marginal transit signal at $T_c = 2459133.4612\ ^{+0.0079}_{-0.0027}$ at a significance of $\sim3\sigma$. The radius ratio is measured to be $R_p/R_s = 0.0244\ ^{+0.0056}_{-0.0071}$.  The posterior transit model and systematic-corrected light curves are shown in the bottom panel of Figure \ref{fig:TOI-1749.03} (a), and the posterior distributions of $T_c$ and $R_p/R_s$ are shown in Figure \ref{fig:TOI-1749.03} (b). Although the significance of the signal is marginal, the measured $T_c$ and $R_p/R_s$ are consistent with the predicted values from TESS, which are shown in blue and red lines in Figure \ref{fig:TOI-1749.03} (b), supporting the detection of this near-threshold planetary candidate in the TESS data (Section \ref{sec:TESS_03}).

\subsection{Validation of the planets}
\label{sec:validation}

We have confirmed the transit signals of all three planetary candidates identified from the TESS data by the ground-based photometric observations (although marginal for TOI-1749b) on target within apertures of 2\farcs6--5\farcs2 radius. The observed transit depths from the ground are all consistent with those from TESS, with no apparent chromaticity for TOI-1749c and TOI-1749d, as shown in Figure \ref{fig:hist_k}. We have also confirmed from the Keck AO observation that there is no companion or background star down to a magnitude difference of 5 within 0\farcs15 from TOI-1749 (Figure \ref{fig:AO}). Therefore the detected transit signals most likely originate from TOI-1749 itself. However, there is still some chance that the each signal comes from an eclipsing binary that contaminates the 0\farcs15 aperture.

We computed false positive probabilities (FPPs) for each planetary candidate using the Python package {\tt vespa} \citep{Morton2015}. The {\tt vespa} package was developed for the statistical validation of planets in bulk, e.g. from the Kepler mission \citep{Morton2016}, which were too numerous or faint to permit detailed follow-up observations sufficient to robustly ascertain their dispositions. {\tt vespa} employs a robust statistical framework to compare the likelihood of the planetary scenario to likelihoods of several astrophysical false positive scenarios involving eclipsing binaries, relying on simulated eclipsing populations based on the {\tt TRILEGAL} Galaxy model \citep{Girardi2005}. As input data to {\tt vespa}, we used the phase-folded TESS light curve for each planet candidate (width $\sim4 \times T_{14}$), the Keck K-band contrast curve (see Figure~\ref{fig:AO}), the 3-$\sigma$ upper limit on the secondary eclipse depth (0.34, 0.28, and 0.46 ppt for TOI-1749b, c, and d, respectively), and a radius of exclusion for the transit signal of 2\farcs6, which corresponds to the minimum aperture radius for ground-based transit observations (see Section \ref{sec:GR_transit_obs}). The FPPs from {\tt vespa} for planets TOI-1749b, TOI-1749c, and TOI-1749d are 0.092, 1.6$\times$10$^{-12}$, and 6.8$\times$10$^{-5}$, respectively. Moreover, since {\tt vespa} does not account for multiplicity, these FPPs are overestimated: \citet{Lissauer2012} demonstrated that a candidate in a system with one (two) or more additional transiting planet candidates is 25 (50) times more likely to be a planet based on multiplicity alone. Application of the 25 (50) multi-boost factor to TOI-1749b yields a probability of 99.6\% (99.8\%) that the candidate is a bona fide planet. Given these considerations, we find that the FPPs of all three candidates fall below the standard validation threshold of 1\%.

\subsection{TTV Analysis}
\label{sec:pdfit}

Because the three planets in this system orbit in a compact region ($\leq$0.07 au), and also the orbits of TOI-1749c and TOI-1749d are close to the 2:1 commensurability, they may exhibit measurable TTVs due to gravitational interactions between the planets. If one assumes that TOI-1749c and TOI-1749d each have a mass of 7 $M_\oplus$ (a mass predicted for a 2.5 $R_\oplus$ planet by \citealt{2017ApJ...834...17C}) and zero free eccentricity, and they are not locked in a MMR, then one can predict that the amplitudes of TTVs are 7.4 and 4.1 minutes for TOI-1749c and d, respectively, with a timescale of the super period of $\sim$610 days \citep{2012ApJ...761..122L}. 

\begin{figure*}[htp!]
    \plotone{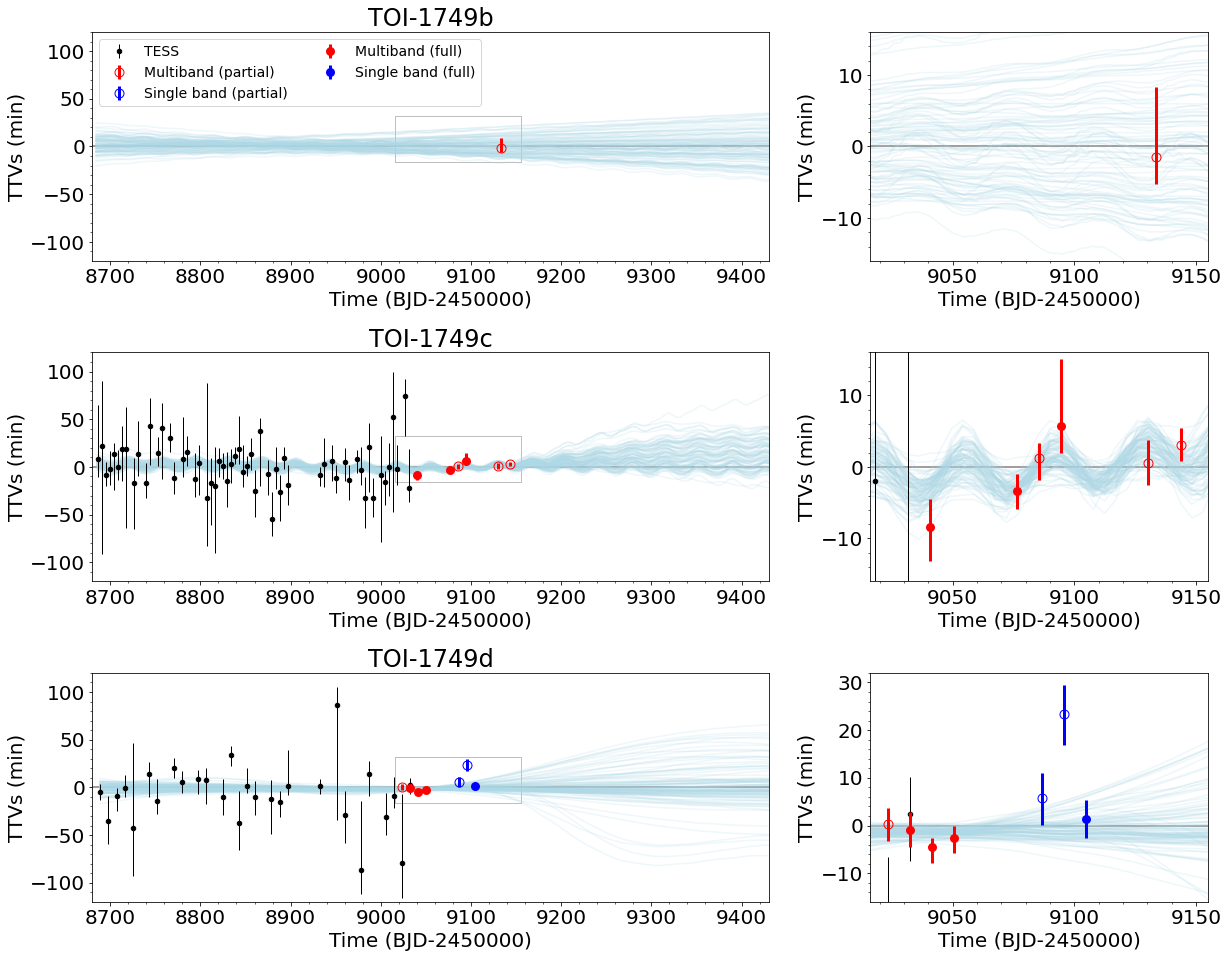}
    \caption{Left panels: transit timing variations of TOI-1749b, TOI-1749c, and TOI-1749d from top to bottom, respectively, with respect to a linear ephemeris calculated from individual $T_c$ measurements from both TESS and ground-based observations. Black, red, and blue data points indicate measurements from TESS, ground-based multiband imagers (MuSCAT2 or MuSCAT3), and ground-based single-band imagers (Sinistro or Spectral), respectively. Each filled or open colored circle derives from ground-based data that cover a full or partial transit, respectively. Light-blue lines represent 100 randomly selected posterior TTV models from the photodynamical analysis. Note that individual $T_c$ values of TOI-1749b were not measured, and are therefore not shown in the plot. Right panels: zoom-in views of the gray boxes on the left.}
    \label{fig:TTVs_3pl}
\end{figure*}

The TESS observation of this target spanned almost one year, which is more than half of the super period. However, as discussed in Section \ref{sec:trfit_TESS}, no evidence of TTVs was found in the TESS data. However, we calculate the orbital periods of TOI-1749c and TOI-1749d based on the individual $T_c$ values from the ground-based observations assuming linear ephemerides to be $4.48934 \pm 0.00011$~days and $9.04535 \pm 0.00032$~days, respectively. These values are both slightly inconsistent with those derived from the TESS data (see Table \ref{tbl:trfit}) by $\sim3\sigma$, which might be a sign of TTV signals. Nevertheless, TTV signals are not apparent even in the ground-based data when we compare them with the linear ephemerides derived using both TESS and ground-based observations (see Figure\ref{fig:TTVs_3pl}), which are  
\begin{eqnarray}
T_{c, {\rm c}} {\rm (BJD_{TDB})} = 2458685.9350\ (26) + 4.489093\ (31) \times E_{\rm c}, \nonumber\\
T_{c, {\rm d}} {\rm (BJD_{TDB})} = 2458688.7790\ (28) + 9.044669\ (76) \times E_{\rm d}, \nonumber
\end{eqnarray}
where $E_{\rm c}$ and $E_{\rm d}$ are transit epochs of TOI-1749c and d, respectively.

Although there are no apparent TTV signals in the measured transit times, TTV signals could be enhanced by the so-called photodynamical modeling \citep{2011Sci...331..562C,2012Sci...337..556C,2015MNRAS.453.2644A}, which directly models the light curves with both transit light-curve and TTV models simultaneously. This method has several advantages over the conventional TTV analysis method. First, this method does not model individual $T_c$'s measured from light curves but instead models the light curves themselves with only the physical parameters of the planetary system. This effectively reduces the total number of free parameters and thus improves the determination of the physical parameters.
Next, with this method the eccentricity ($e$) and longitude of periastron ($\varpi$) can be shared between the TTV and transit light-curve models. Because $a/R_s$ can be converted from $e$, $\varpi$, $P$, and stellar density $\rho_s$, the constraints on $e$ and $\varpi$ from the TTV model allow one to use $\rho_s$ as a common parameter within multiple planets in the same system instead of $a/R_s$ for each planet in the transit light-curve model; this also helps to reduce the number of free parameters.
Finally, this technique allows one to properly estimate the uncertainties of the parameters even if individual $T_c$'s have asymmetric (non-Gaussian) posterior distributions due to, e.g., partial transit coverage or systematics in the light curve.

To constrain the mass and eccentricity of the planets, and obtain better estimates of other transit parameters, we performed a joint analysis of the TESS and ground-based (those check-marked in Table \ref{tbl:observing_log}) light curves by photodynamical modeling. In this analysis, we used the detrended TESS light curves, in which the systematics in the original light curves are corrected by the systematic model derived from the individual $T_c$ fit in Section \ref{sec:trfit_TESS}, while we used the un-detrended light curves (same as the ones used in Section \ref{sec:trfit_GR_01_02}) for the ground-based data, in which the systematic and photodynamical models could be non-negligibly correlated due to the similar timescales between the systematics and transits. We calculated TTV models using {\tt TTVFast} \citep{2014ApJ...787..132D}, where we used as the parameters the mean orbital period $P$, planet-to-star mass ratio $M_p/M_s$, eccentricity and argument of periastron in the forms of $\sqrt{e} \sin\varpi$ and $\sqrt{e} \cos\varpi$, and orbital phase $\varpi + {\mathcal M}$, where ${\mathcal M}$ is the mean anomaly at a reference time (BJD$=2458683.354296$). We assumed coplanar orbits with an orbital inclination of 90$^\circ$ and argument of ascending node of 180$^\circ$ for all planets\rev{, and used a time step of 0.05 days for the dynamical calculation.} For the transit model, we applied the same model as described in Section \ref{sec:trfit_TESS} and Section \ref{sec:trfit_GR_01_02}, but dropped the assumption of circular orbits and allowed nonzero values for $e$ and $\varpi$, which were shared with the TTV model. The $T_c$ of individual transits were calculated by the TTV model. We assumed an achromatic transit depth and used a single $R_p/R_s$ for all bands for each planet. For the systematic models of ground-based data, we fixed $c_x$ and $c_y$ to the values determined in Section \ref{sec:trfit_GR_01_02} to reduce the number of free parameters, while letting $\rho$ and $\sigma$ be free in the same way as in Section \ref{sec:trfit_GR_01_02}. We also fixed the white-noise jitter values to those determined in Sections \ref{sec:trfit_TESS} and \ref{sec:trfit_GR_01_02} for TESS and ground-based data, respectively. The total number of free parameters for three planets was 23.

We then estimated the posterior distributions of the parameters in an MCMC analysis using {\tt emcee}.
Because the likelihood function with respect to the parameters of a TTV model tends to be multimodal, it is not efficient to start MCMC chains from one single location in the parameter space. Alternatively, we ran MCMC with 400 walkers in which $M_p$, $e$, and $\varpi$ were initialized with uniformly random values in the ranges of [1,20] $M_\oplus$, [0.01, 0.2], and [$-\pi, \pi$], respectively, while the other parameters were initialized around the median values of the posteriors calculated in Section \ref{sec:trfit_TESS}. In addition, we sampled MCMC chains with the differential evolution algorithm, which can efficiently find global maxima of the likelihood function even if it is multimodal. After running MCMC for $4\times10^4$ steps, we sorted the walkers by likelihood of the last step and discarded the lower half of the walkers, because many of them were stuck at local minima of the likelihood function. We then continued MCMC runs with the remaining walkers for a sufficient number of steps until the chains converged, and sampled the last $8\times10^4$ steps, which span intervals roughly 20 times the auto-correlation lengths, to construct the posterior probability distributions of the parameters.

We summarize the median values and 1$\sigma$ confidence intervals of the parameters in Table \ref{tbl:pdfit_model}. In Table \ref{tbl:pdfit_derived} we also show the derived parameters of $e$, $\varpi$, $R_p$, and $M_p$, where we apply $R_s$ and $M_s$ derived in Section \ref{sec:host_star} to calculate $R_p$, and $M_p$, respectively. In Figure \ref{fig:TTVs_3pl}, we display randomly selected 100 posterior TTV models for each planet along with the individually measured transit times in the previous sections for comparison.

From this analysis, we find that a nonzero mass is preferred for all planets at the $\sim$1~$\sigma$ level ($19^{+22}_{-13}$ $M_\oplus$, $2.1^{+5.7}_{-1.6}$ $M_\oplus$, and $4.3^{+6.2}_{-3.5}$ $M_\oplus$ for TOI-1749b, c, and d, respectively), which is indicative of TTV signals in the data. To further check for the significance of the TTV signals, we also fit a constant period model to the same datasets assuming all planets have circular orbits, and compared the minimum $\chi^2$ values with that of the photodynamical model. As a result, we find that the $\chi^2$ value of the photodynamical model improves over the null-TTV model by 762. Given the total numbers of data points relevant to individual transits of $\sim$45,000 ($\sim25,000$ from the ground-based data) and the number of additional free parameters of 9, this $\chi^2$ improvement corresponds to a BIC improvement of $\sim$666, which indicates that the TTV model better describes the observed data over the null-TTV one. Note that this $\chi^2$ improvement entirely comes from the ground-based data. The portion of $\chi^2$ from the TESS data increased by 108, which may indicate the presence of uncorrected systematics in the TESS data, uncorrected systematics in the ground-based data that generates a bias in the TTV model in the region of the TESS data, or an additional body in the system. Further TTV observations are required to investigate these possibilities.

While the planetary masses are not detected at high significance, we place strong upper limits on the mass of 57~$M_\oplus$, 14~$M_\oplus$, and 15~$M_\oplus$ (95\% confidence level) for TOI-1749b, c, and d, respectively, confirming that all three planets have a mass well within the planetary range ($< 13 M_{\rm Jup} = 4132 M_\oplus$).

\begin{deluxetable*}{lccccc}[htp!]
\label{tbl:pdfit_model}
\tablecaption{Results of photodynamical analysis.}
\tablehead{
\colhead{Parameter} & \colhead{Unit} & \colhead{Host star} & \colhead{TOI-1749b}
& \colhead{TOI-1749c} & \colhead{TOI-1749d}
}
\startdata
$\rho_s$ & g cm$^{-3}$ & $5.02\ ^{+0.59}_{-0.46}$ & ... & ... &...\\
$u_1$ (TESS) && $0.329\ ^{+0.015}_{-0.013}$ &...&...&...\\
$u_2$ (TESS) && $0.249 \pm 0.019$ &...&...&...\\
$P$ & days &... & $2.38839\ ^{+0.00031}_{-0.00066}$ & $4.4929\ ^{+0.0038}_{-0.0027}$ & $9.0497\ ^{+0.0049}_{-0.0032}$\\
$b$ &&... & $0.75\ ^{+0.11}_{-0.20}$ & $0.36\ ^{+0.09}_{-0.31}$ & $0.717\ ^{+0.028}_{-0.032}$\\
$R_p/R_s$&&... & $0.0232\ ^{+0.0032}_{-0.0029}$ & $0.0353 \pm 0.0008$ & $0.0421 \pm 0.0010$\\
$M_p/M_s$& $10^{-5}$ &... & $10\ ^{+11}_{-7}$ & $1.1\ ^{+2.9}_{-0.8}$ & $2.2\ ^{+3.2}_{-1.8}$\\
$\sqrt{e} \cos \varpi$ &&... & $-0.05 \pm 0.12$ & $0.025\ ^{+0.059}_{-0.043}$ & $-0.02 \pm 0.08$\\
$\sqrt{e} \sin \varpi$ &&... & $0.08 \pm 0.13$ & $0.03\ ^{+0.09}_{-0.06}$ & $-0.03 \pm 0.11$\\
$\varpi+{\mathcal M}$ & radians &... & $2.051\ ^{+0.035}_{-0.069}$ & $1.102 \pm 0.010$ & $0.943^{+0.019}_{-0.022}$\\
\hline
\enddata
\end{deluxetable*}

\begin{deluxetable*}{lcccc}[htp!]
\label{tbl:pdfit_derived}
\tablecaption{Derived parameters from the results of photodynamical analysis.}
\tablehead{
\colhead{Parameter} & \colhead{Unit} & \colhead{TOI-1749b} & \colhead{TOI-1749c} & \colhead{TOI-1749d}
}
\startdata
Planetary radius ($R_p$) & R$_\oplus$ & $1.39\ ^{+0.21}_{-0.19}$ & $2.12 \pm 0.12$ & $2.52 \pm 0.15$\\
Planetary mass ($M_p$) & M$_\oplus$ & $19\ ^{+22}_{-13}$ (57)  & $2.1\ ^{+5.7}_{-1.6}$ (14) & $4.3\ ^{+6.2}_{-3.5}$ (15)\\
Eccentricity ($e$) && $0.02\ ^{+0.06}_{-0.02}$ (0.14) & $0.007\ ^{+0.007}_{-0.005}$ (0.019) & $0.015\ ^{+0.017}_{-0.011}$ (0.062)\\
Longitude of periastron ($\varpi$) & radians & $2.3\ ^{+1.7}_{-1.0}$ & $1.8\ ^{+3.2}_{-1.0}$ & $3.8\ ^{+1.3}_{-2.2}$\\
Orbital inclination ($i$) & degrees & $86.4\ ^{+0.9}_{-0.6}$ & $88.8\ ^{+1.0}_{-0.3}$ & $88.53\ ^{+0.11}_{-0.09}$\\
Semi-major axis ($a$) & au & $0.0291 \pm 0.0005$ & $0.0443 \pm 0.0008$ & $0.0707 \pm 0.0012$\\
Equilibrium temperature ($T_{\rm eq}$) \tablenotemark{a}& K & $831 \pm 18$ & $673 \pm 15$ & $533 \pm 12$\\
\enddata
\tablecomments{The value in parentheses represents the 95\% confidence upper limit.}
\tablenotetext{a}{Calculated for a planet with zero albedo and a constant surface temperature.}
\end{deluxetable*}

\section{Discussion}
\label{sec:discussion}

\subsection{Stability of the system}
We carried out a set of dynamical simulations to study the long-term stability of the system.
We took the parameters reported in Tables~\ref{tbl:pdfit_model} and \ref{tbl:pdfit_derived} and drew 60,000 samples from the parameters posteriors as initial parameters for the dynamical simulation. Each parameter set was integrated for 10$^{9}$ orbits of the inner planet orbital period using the Stability of Planetary Orbital Configurations Klassifier \citep[\texttt{SPOCK};][]{spock_2020}. The system was found to be dynamically stable for the whole parameter posterior space with a median value of the spock stability probability of 0.84.

\begin{figure*}[htp!]
    \plotone{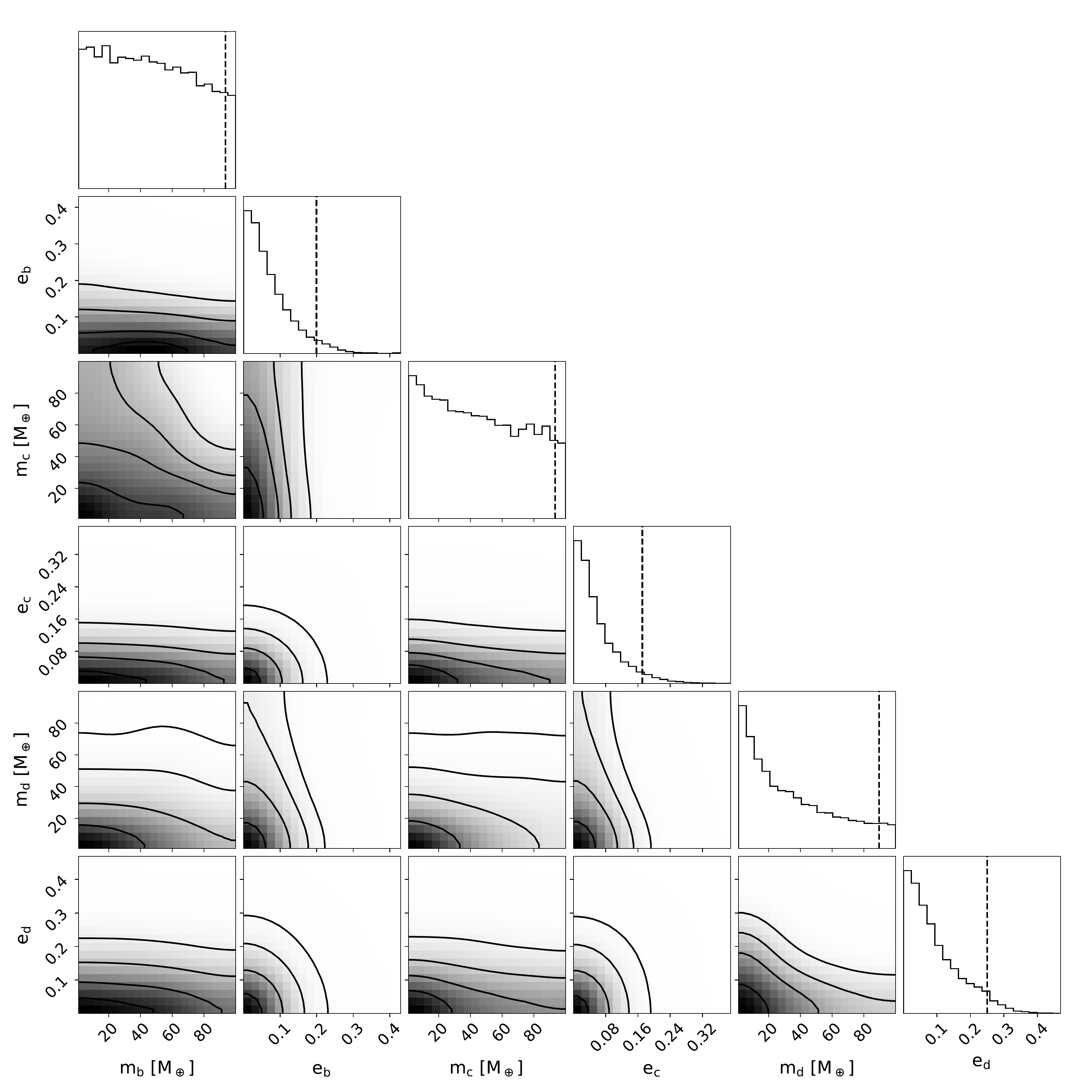}
    \caption{\rev{Corner plot of the posterior distributions for the planet masses and orbital eccentricities for TOI-1749b, c, and d from the stability analysis. The contours indicate 1, 2, 3, and 4$\sigma$ confidence levels from inside to outside, and the dashed lines mark the 95\% confidence interval.}}
    \label{fig:spock_mcm}
\end{figure*}

Since the photodynamical analysis in Section~\ref{sec:pdfit} gave only upper limits (95\% confidence level) on the masses for TOI-1749b, c, and d, we employed a set of dynamical simulations as described above combined with an MCMC sampling using {\tt emcee} to explore the overall stability of the system.  We sampled the \texttt{SPOCK} stability probability using 50 walkers, running the MCMC over 15,000 iterations of which we used the last 10$^7$ and used the first $5 \times 10^6$ as burn-in. We allowed for eccentricities up to 0.5 and planet masses up to 100 $M_\oplus$. The posteriors for these two parameters are show in Figure~\ref{fig:spock_mcm} where we report the 95\% confidence level as the upper limits. Although the eccentricities have upper limits of 0.19, 0.17, and 0.25, for TOI-1749b, c, and d, respectively, the planet masses for the innermost planets are not constrained at all and only marginally constrained for the outer planet. 
Those simulations show that the adopted solution (Table~\ref{tbl:pdfit_model}) lies within the stable regions.

\subsection{Possible compositions of the planets}

\begin{figure}[htp!]
    \centering
    \includegraphics[width=8.6cm]{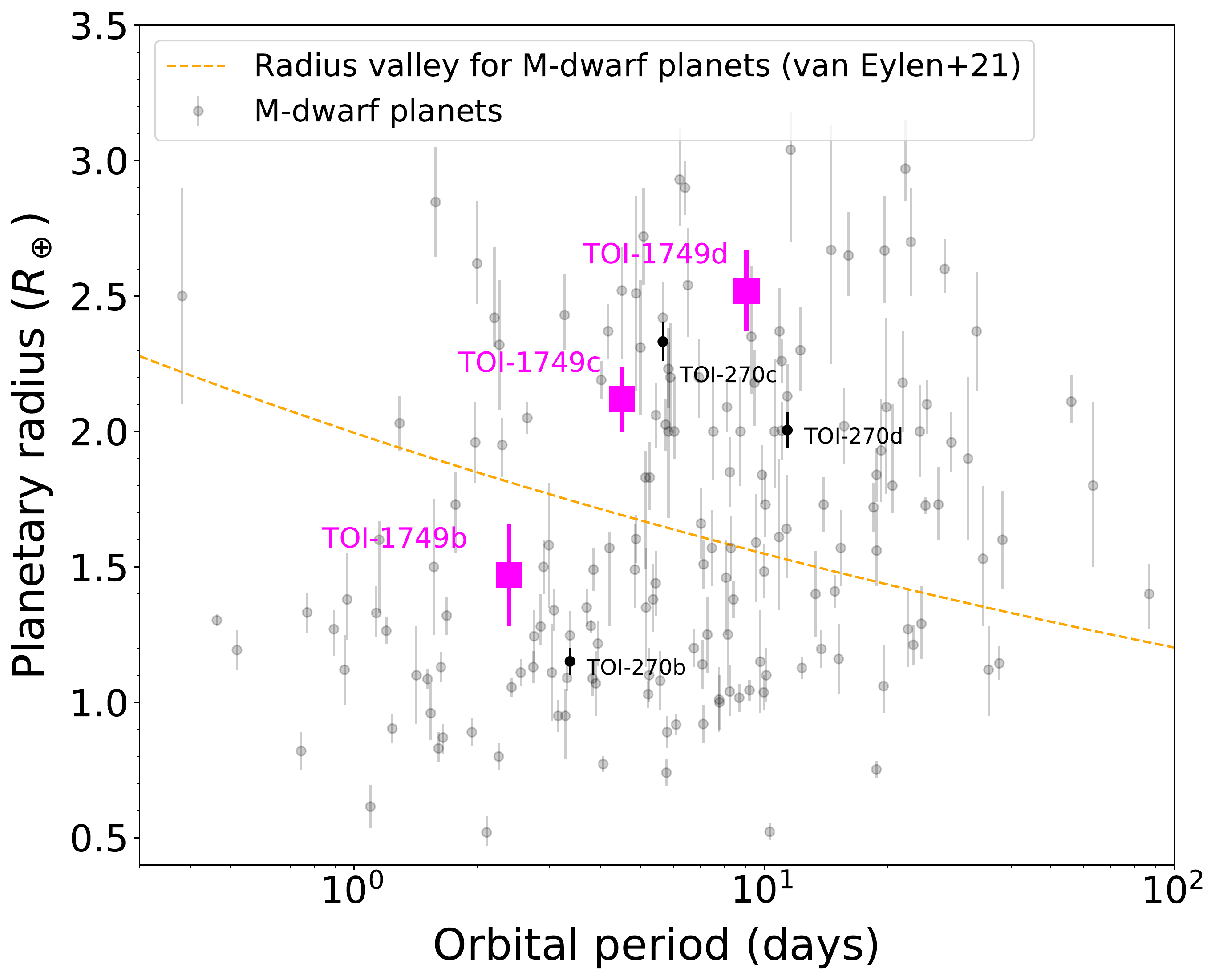}
    \caption{Period and radius diagram for planets around M dwarfs. The gray dots are known planets around M dwarfs taken from the NASA Exoplanet Archive. The magenta squares and black dots are for planets in the TOI-1749 and TOI-270 systems, respectively. The orange dashed line indicates the location of the proposed radius valley for M dwarf planets by \cite{2021MNRAS.tmp.2193V}. }
    \label{fig:P_vs_Rp}
\end{figure}

In the period-radius plane, the innermost planet (TOI-1749b) and the outer two planets (TOI-1749c and d) are located below and above, respectively, the proposed location of the radius valley for planets around M dwarfs \citep{2021MNRAS.tmp.2193V} as shown in Figure \ref{fig:P_vs_Rp}. Several studies have shown that this radius valley can be a consequence of photoevaporation and/or core-powered loss of hydrogen envelopes on top of rocky cores; the hydrogen envelopes of planets smaller than a certain size for a given orbital period are selectively dissipated \citep[e.g.,][]{Owen_2017,2018MNRAS.476..759G}. Considering the fact that the three planets have similar orbits in the same system, they could have similar initial compositions. If so, then their current radii can be explained by a scenario wherein they all initially consisted of a rocky core surrounded by a thin hydrogen envelope, and then only the innermost planet had lost its envelope due to the photoevaporation and/or core-powered mass loss mechanisms. 
\rev{For comparison, we also show the locations of TOI-270b, c, and d in Figure \ref{fig:P_vs_Rp}, which are a benchmark triple orbiting the nearby M dwarf TOI-270 \citep[also known as L231-32;][]{2019NatAs...3.1099G}. This trio also has the same trend that the innermost and the outer two planets are located below and above the predicted radius valley, respectively.}

\begin{figure*}[htp!]
    \centering
    \includegraphics[width=10cm]{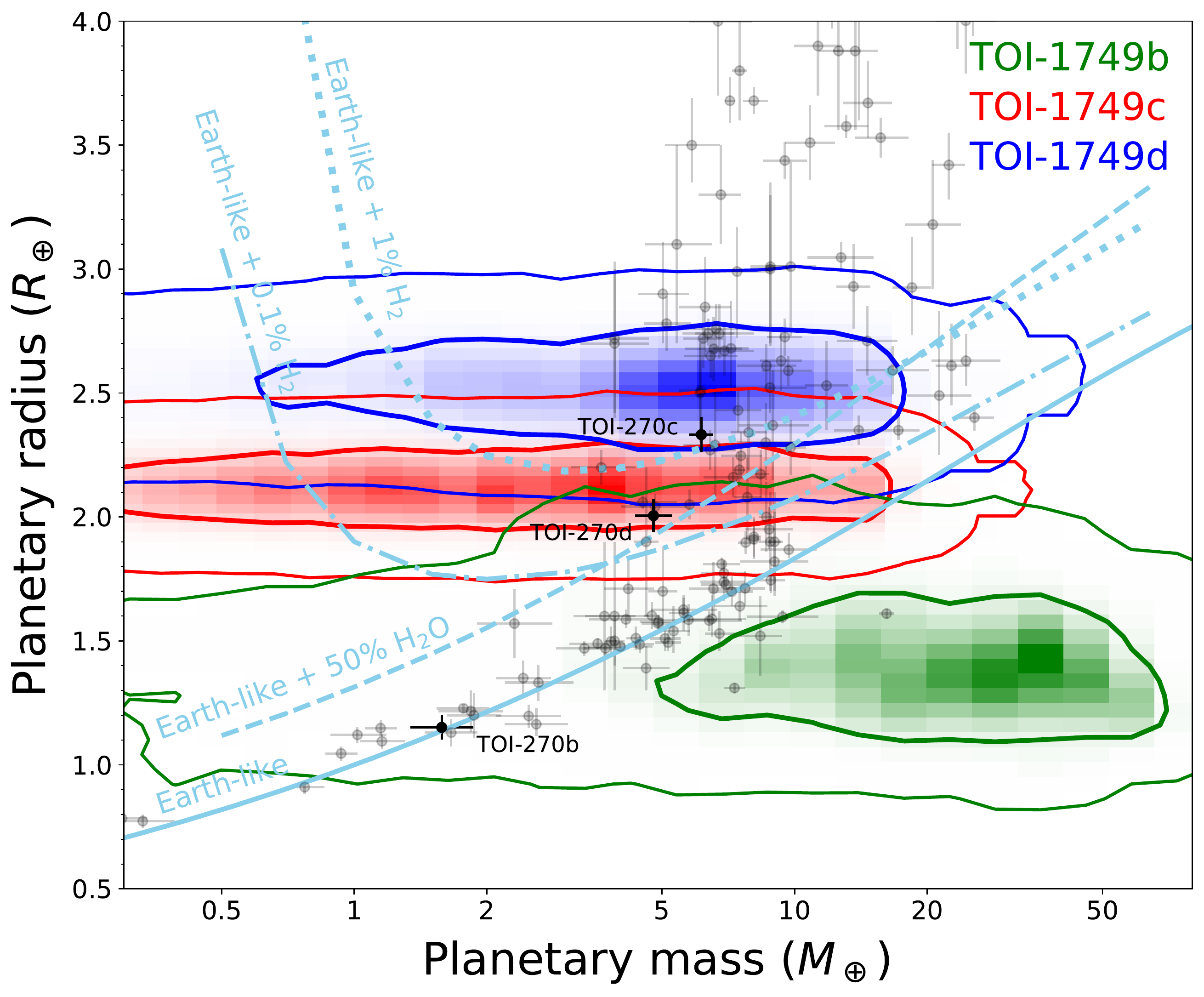}\\
    \caption{\label{fig:Mp_vs_Rp}
    \rev{Mass} and radius diagram. The green, blue, and red color maps show the densities of posterior probabilities from the photodynamical analysis for TOI-1749b, TOI-1749c, and TOI-1749d, respectively. For each planet, the inner thick and outer solid contours indicate 1$\sigma$ and 2$\sigma$ confidence intervals, respectively. Gray dots with error bars are known exoplanets with mass and radius determined with $<$20\% uncertainties, taken from the NASA Exoplanet Archive. The light-blue solid, dashed, dash-dotted, and dotted lines are theoretical mass-radius relations from \cite{2019PNAS..116.9723Z} for planets with an Earth-like rocky composition (32.5wt\% Fe + 67.5wt\% MgSiO$_3$), a rocky+icy composition (50wt\% Earth-like rocky core + 50wt\% H$_2$O), an Earth-like rocky core + 0.1wt\% hydrogen envelope, and an Earth-like rocky core + 1wt\% hydrogen envelope, respectively, where 500~K is assumed for the planetary equilibrium temperature.}
\end{figure*}

In the top panel of Figure \ref{fig:Mp_vs_Rp}, we show the densities of the posterior probabilities of the mass and radius of TOI-1749b (green), TOI-1749c (blue), and TOI-1749d (green) derived from the photodynamical analysis, along with the known planets with mass and radius both measured taken from the NASA Exoplanet Archive\footnote{\url{https://exoplanetarchive.ipac.caltech.edu/}}. We also show theoretical mass-radius relations from \cite{2019PNAS..116.9723Z} for planets with an Earth-like rocky composition (32.5wt\% Fe + 67.5wt\% MgSiO$_3$), a rocky+icy composition (50wt\% Earth-like rocky core + 50wt\% H$_2$O), an Earth-like rocky core + 0.1wt\% hydrogen envelope, and an Earth-like rocky core + 1wt\% hydrogen envelope (light-blue solid, dashed, dashed-dotted, and dotted lines, respectively). Although the uncertainties in the masses of the TOI-1749 planets are large enough for the masses and radii to be consistent with a range of compositions, \rev{the posterior probabilities of} the innermost (TOI-1749b) and the outer two (TOI-1749c and d) \rev{planets are the most dense} below and above the theoretical line for an Earth-like composition, respectively. This result is consistent with the above scenario, in which TOI-1749b currently consists of a bare rocky core, while TOI-1749c and d still have a hydrogen envelope on top of a rocky core.
\rev{We note that} the median value of the mass of TOI-1749b (19 $M_\oplus$) is unusually large for its size ($\sim1.4 R_\oplus$)\rev{, which, however, } 
could be biased by any systematics in the data, given the low significance of the TTV signals. In particular, there is a well-known mass-eccentricity degeneracy in TTV models, and \rev{a smaller mass and larger eccentricity of TOI-1749b could also explain the data with only a slightly lower likelihood. Additional observations are required to test for this possibility.}

We note that\rev{, thanks to its bright host star,} the masses and radii of \rev{the benchmark trio of the TOI-270 system have been precisely measured through radial velocity and TESS photometry, respectively \citep{2021MNRAS.tmp.2193V}, from which the innermost and the outer two planets of this system were confirmed to be consistent with a rocky and rocky+hydrogen envelope compositions, respectively (see Figure \ref{fig:Mp_vs_Rp}).}

\subsection{Period ratio of TOI-1749c and d}
\label{sec:period_ratio}

One interesting feature of this system is that the planetary pair of TOI-1749c and d has a period ratio of 2.015, which is very close to the exact 2:1 commensurability with an outward departure of only 0.7\%. This feature is shared with two other pairs of planets around M dwarfs,  TOI-270 c and d and TOI-175 (also known as L98-59) c and d \citep{2019AJ....158...32K,2019A&A...629A.111C}, which have period ratios of 2.011 and 2.019, respectively.
These specific period ratios are {\it a priori} less likely to be of primordial origin than the consequence of resonant capture by convergent migration in protoplanetary disks, which might be followed by some resonant repulsion effects.

\begin{figure}[htp!]
    \centering
    \includegraphics[width=8cm]{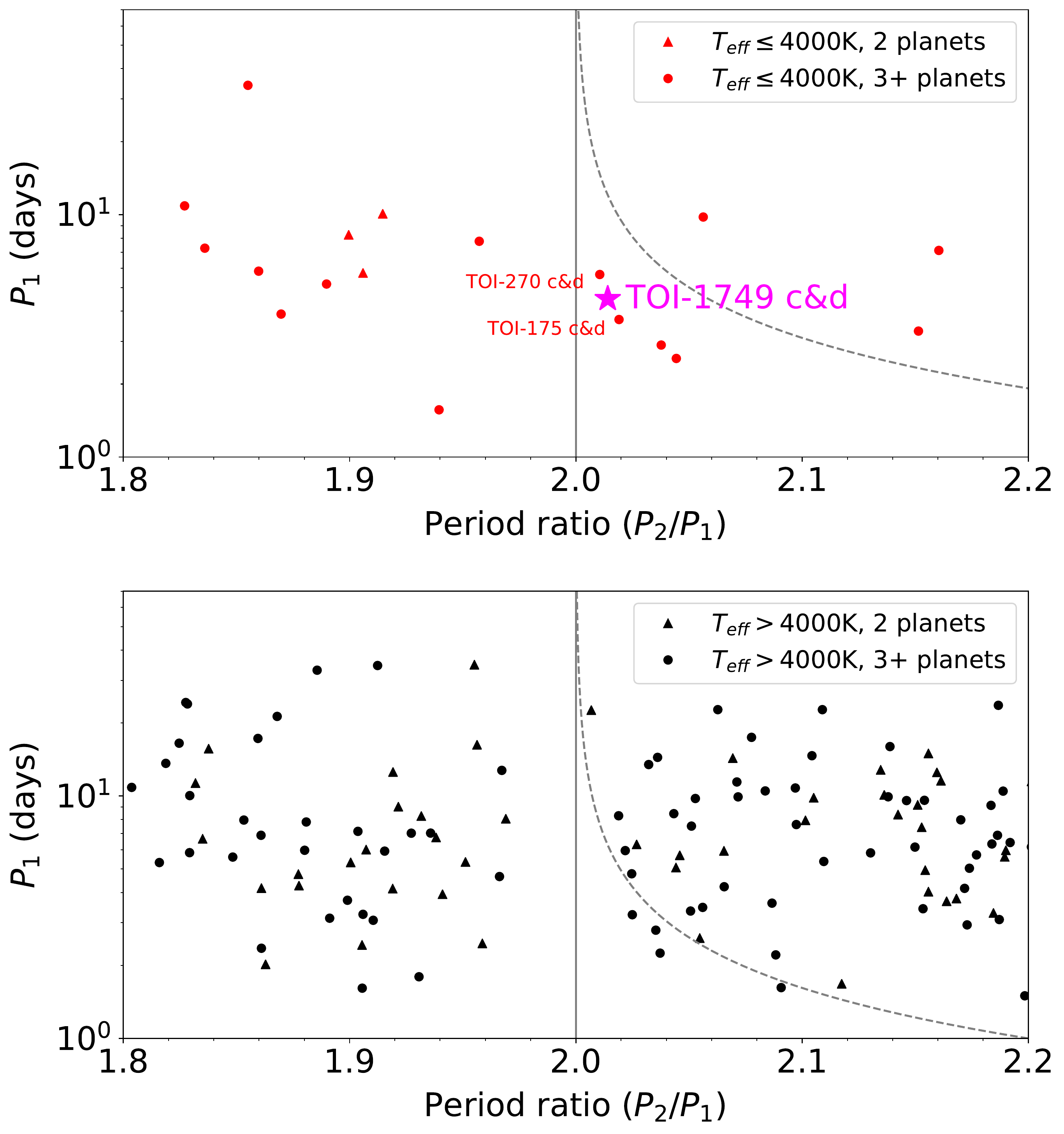}
    \caption{Orbital period of inner planets and period ratio of known transiting planetary pairs close to the 2:1 commensurability (gray vertical lines) that orbit M dwarfs (top; $T_{\rm eff} \leq 4000$~K) and FGK dwarfs (bottom; $T_{\rm eff} > 4000$~K). Triangles and circles are planetary pairs in planetary systems in which no other planet has so far been discovered and at least one additional planet has been discovered, respectively. 
    The dashed lines indicate the distance from the exact 2:1 commensurability to which a planetary pair is expected to be repelled by the tidal dumping effect in 5 Gyr, assuming that the inner planet has $R_p = 2 R_\oplus$, $M_p = 10 M_\oplus$, and $Q' \equiv 3Q/2k_2 = 150$. The stellar mass is assumed to be 0.5 $M_\odot$ and 1 $M_\odot$ for the top and bottom panels, respectively.}
    \label{fig:Pratio1_vs_P1}
\end{figure}

A period ratio just outside of a commensurability has also been commonly observed in planetary pairs around FGK stars, which were mainly discovered by Kepler \citep{2011ApJS..197....8L}. However, in the case of the planetary pairs around FGK stars, it is relatively rare that the period ratio of a pair departs from the exact 2:1 commensurability by only 1\% or less, as shown in Figure \ref{fig:Pratio1_vs_P1}. In this figure, the period ratio of planetary pairs with radii not larger than 4 $R_\oplus$ are shown as a function of orbital period of the inner planet for M dwarf and FGK dwarf systems in the top and bottom panels, respectively. This possible difference between M dwarf and FGK dwarf systems implies that the dominant mechanisms that repel planetary pairs around FGK dwarfs from exact commensurabilities would not work in the same way for those around M dwarfs, or at least for the above three pairs.

So far, various mechanisms have been proposed to explain the paucity and pileup of planetary pairs at and just outside the exact commensurabilities, respectively, observed in the Kepler-discovered multiplanet systems. These include tidal interactions between the planets and central star \citep{2012ApJ...756L..11L,2013AJ....145....1B,2014A&A...570L...7D}, planet-disk interactions \citep{2013ApJ...778....7B}, planet-planetecimal interactions \citep{2015ApJ...803...33C}, and dynamical instabilities of planets in resonant chains \citep{2017MNRAS.470.1750I}. 

One of the most frequently discussed mechanisms among them has been tidal interaction.
In this scenario, the forced eccentricities of two near-resonant planets raised by the resonance are damped by the tidal effect, which works more efficiently for the inner planet, leading to a repulsion from the resonance. Using Equation (26) of \cite{2012ApJ...756L..11L}, one can express the fractional distance of the period ratio from the exact 2:1 commensurability that a planetary pair with the initial period ratio of 2 obtains by the tidal effect after the time $t$ as

\begin{eqnarray}
\label{eq:delta_2to1}
\Delta_{\rm 2:1} &\sim& 0.01
\left( \frac{Q'_1}{150}\right)^{-1/3}
\left( \frac{M_{p,1}}{10 M_\oplus}\right)^{1/3} 
\left( \frac{R_{p,1}}{2 R_\oplus}\right)^{5/3} \nonumber\\
&\times& \left( \frac{M_*}{M_\odot}\right)^{-8/3}
\left( \frac{P_1}{5 {\rm days}}\right)^{-13/9}
\left( \frac{t}{5 {\rm Gyr}}\right)^{1/3},
\end{eqnarray}
where $Q' \equiv 3Q/2k_2$, $Q$ is tidal quality factor, $k_2$ is tidal Love number, $M_p$ is planetary mass, $R_p$ is planetary radius, $P$ is orbital period, $M_*$ is stellar mass, and the subscript ``1'' denotes the inner planet. This equation indicates that the resonant repulsion effect has a strong, negative dependence on the stellar mass, making a factor of $\sim$6 difference in $\Delta_{2:1}$ between M ($M_*=0.5 M_\odot$) and G ($M_*=1 M_\odot$) dwarf systems for a given set of the other parameters, as shown by dashed lines in Figure \ref{fig:Pratio1_vs_P1}. Note that the other parameters in Equation (\ref{eq:delta_2to1}), i.e., $M_p$, $R_p$, $Q'$, and $t$, on which $\Delta_{\rm 2:1}$ has weaker dependence than on $M_*$, could also be somewhat different between the M and FGK samples. The important point here is that despite the strong and negative dependence of $\Delta_{2:1}$ on $M_*$, there do exist planetary pairs closer to the exact 2:1 commensurability around M dwarfs than any of those around FGK dwarfs for a given orbital period of the inner planet. This fact implies that there are some mechanisms that counteract the tidal repulsion effect selectively in the M dwarf systems, or alternative resonant repulsion mechanisms that do not depend on (or have positive dependence on) the stellar mass play a major role in both types of systems.

\begin{figure*}[htp!]
    \centering
    \includegraphics[width=14cm]{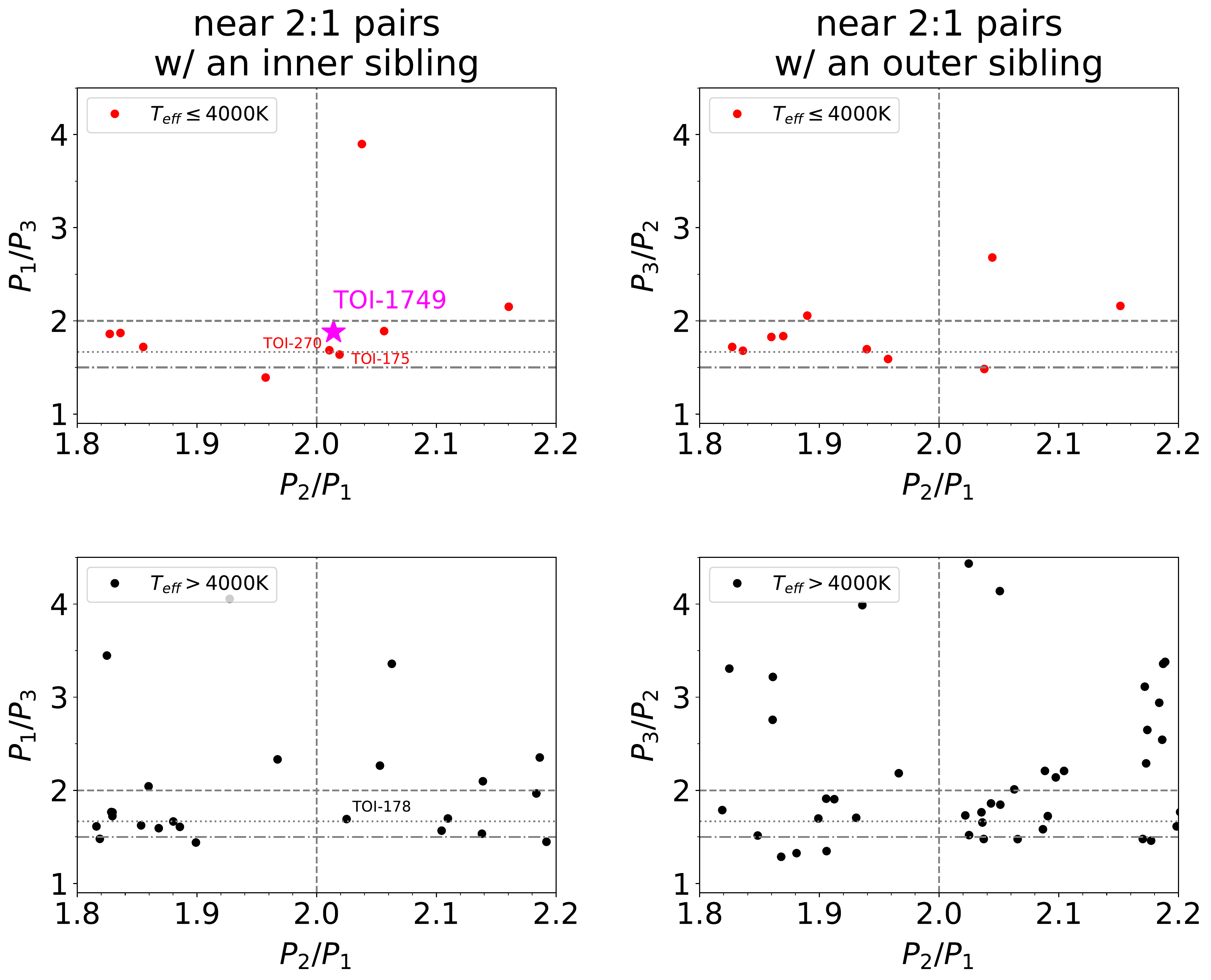}
    \caption{Period ratios of planetary trios including a near 2:1 commensurability pair. $P_1$, $P_2$, and $P_3$ denote the orbital period of the inner planet in a near 2:1 commensurability pair, that of the outer planet in the same pair, and that of another planet inside (left panels) or outside (right panels) of the near 2:1 commensurability pair, respectively. Top and bottom panels show planetary trios around M dwarfs ($T_{\rm eff} \leq 4000$~K) and FGK dwarfs ($T_{\rm eff} > 4000$~K), respectively. The gray dashed, dash-dotted, and dotted lines indicate the locations of the exact 2:1, 2:3, and 3:5 commensurabilities, respectively. This figure shows that additional planets to planetary pairs near 2:1 commensurability around M (FGK) dwarfs tend to have an inside (outside) orbit. Note that the TOI-178 system shows an opposite property; the host star however has an effective temperature that is close to the boundary \citep[$T_{\rm eff}=4316$K,][]{2021A&A...649A..26L}.}
    \label{fig:Pratio1_vs_Pratio2}
\end{figure*}

Another common feature among the near 2:1 planetary pairs in the TOI-175, TOI-270, and TOI-1749 systems is that they all have an additional inner planet. The period ratios of the inner adjacent pairs (i.e., the period ratio of the middle to the innermost planets) are all between 1.5 and 2, as shown in the top left panel of Figure \ref{fig:Pratio1_vs_Pratio2}. However, near 2:1 period-ratio planetary pairs in FGK systems tend to have an additional planet at an outer orbit rather than inner one, as shown in the bottom panels of Figure \ref{fig:Pratio1_vs_Pratio2}. Note that the TOI-178 system is an exception to this trend; i.e., the near 2:1 period ratio pair of TOI-178c and d has an additional inner planet (TOI-178b) despite the fact that the host star is a K dwarf. However, the host star has an effective temperature of $T_{\rm eff}=4316$~K \citep{2021A&A...649A..26L}, which is close to the temperature range of M dwarfs $<4000$~K. 

Several previous works have pointed out that planetary pairs in high multiplicity ($>$2 planets) systems tend to avoid resonant repulsion \citep{2012ApJ...756L..11L,2015MNRAS.448.1956S}, probably due to extra dynamical forces caused by the additional planets. The different tendency in the orbital architecture of planetary trios between M and FGK dwarf systems, seen in Figure \ref{fig:Pratio1_vs_Pratio2}, might suggest that the specific orbital configuration of the TOI-175, TOI-270, and TOI-1749 systems has played a role in competing with the resonant repulsion effects. Given that TOI-1749 is the third such example, planetary trios in which the outer pair has a period ratio just beyond 2 might be relatively common around M dwarfs compared to FGK dwarfs, while a rigorous statistical test is beyond the scope of this paper. If confirmed, this fact might reflect the planetary formation and migration processes, which should depend on the stellar mass, or more specifically, the properties of the protoplanetary disks.

\subsection{Prospects for future follow-up observations}

Given that TOI-1749 is one of a few M dwarf systems that host multiple transiting planets including a pair close to the 2:1 MMR, further follow-up observations are encouraged.

Continuous monitoring of TTVs is of particular importance in confirming and narrowing down the masses and eccentricities of the planets. As demonstrated in this work, although the TTV amplitudes in the outer two planets are too small for TESS to detect, ground-based 1-2m class telescopes can measure the times of individual transits with a high enough precision. Multiband cameras like MuSCATs are particularly efficient for this purpose. As seen in Figure \ref{fig:TTVs_3pl}, the timing precisions achieved for TOI-1749d from MuSCAT2 (red points) are better than any other data from single-band imagers (blue ones), regardless of whether the transit coverage is full or not. In addition, detecting transits of TOI-1749b would be challenging for single-band imagers mounted on 1-2m or smaller class telescopes.

Radial velocity (RV) observations can also yield independent confirmation of the planetary masses. The expected RV amplitudes from the individual planets are $\sim2$--3 m~s$^{-1}$ each if they have a mass of $\sim$3--4 $M_\oplus$, which is within reach of the current world-best instruments. One drawback is the faintness of the host star ($V=13.9$, $J=11.1$), which would require instruments on large aperture ($\sim$8~m) telescopes such as Subaru/IRD and Gemini/Maroon-X.

This system may not be a prime target for planetary atmospheric study due to the relative faintness of the host star. Using Equation (1) of \cite{2018PASP..130k4401K}, we estimate the transmission spectroscopy metric (TSM) of TOI-1749c and TOI-1749d both to be 32, where we apply the masses estimated using the empirical relation of $M_p = 1.436 R_p^{1.70} M_\oplus$ from \citet{2017ApJ...834...17C} (5.1 and 6.9 $M_\oplus$ for TOI-1749c and d, respectively). These values are a factor of 3--4 smaller than those for their analogs of TOI-270c and TOI-270d \citep{2020MNRAS.495..962C,2021MNRAS.tmp.2193V}. However, because TSM is inversely proportional to the planetary mass, they would become good atmospheric targets if the true mass is unusually small, such as 1--2 $M_\oplus$, which is still allowed by the current observations (see Figure \ref{fig:Mp_vs_Rp}). We note that TOI-1749 is located very close to the northern continuous viewing zone of the James Webb Space Telescope, allowing flexible scheduling with the promising new space telescope.

\section{Summary}
\label{sec:summary}

We have detected one super-Earth- and two sub-Neptune-sized planetary candidates around the M dwarf TOI-1749 from TESS transit photometry, and validated their planetary nature from ground-based low-resolution spectroscopy, high-resolution imaging, and multiband transit photometry.
In addition, using photodynamical models we have been able to place 95\% confidence level upper limits on the masses of TOI-1749b, TOI-1749c, and TOI-1749d of 57, 14, and 15 $M_\oplus$, respectively. The radii of the innermost planet (TOI-1749b) and the outer two planets (TOI-1749c and d) are found to be at the lower and upper sides of the proposed radius valley, respectively, being consistent with the scenario that only the envelope of the innermost planet had been stripped away by photoevapolation and/or core-powered mass loss. The tentatively constrained masses of the planets from photodynamical modeling are consistent with this scenario; i.e., that the compositions of the innermost planet and the outer two planets can be explained by a bare rocky core and a rocky core + thin hydrogen envelope, respectively. We have confirmed that the system is dynamically stable \rev{for at least 10$^9$ orbits of the innermost planet.}

The outer planetary pair has a period ratio very close to the 2:1 commensurability (2.015), sharing the orbital architecture with the other M dwarf systems TOI-270 and TOI-175. This characteristic architecture might be a consequence of common planetary formation and migration processes in these systems. Further follow-up observations of this system would be worth pursuing to characterize the system in more detail. Mass determination by TTVs is of particular importance in confirming the planetary compositions, for which ground-based multiband instruments like the MuSCATs are efficient as demonstrated by this work.

\begin{acknowledgements}

We acknowledge Masahiro Ikoma for supporting the development of MuSCAT3 behind the scenes. 
We acknowledge Sudhish Chimaladinne, Srihan Kotnana, David Vermilion, Deven Combs, Kevin Collins, and Peter Plavchan for observations and analyses using the GMO telescope.
A.F. thanks Sho Shibata and Tadahiro Kimura for useful discussions on possible formation scenarios of the TOI-1749 system.

Funding for the TESS mission is provided by NASA's Science Mission Directorate. We acknowledge the use of public TESS data from pipelines at the TESS Science Office and at the TESS Science Processing Operations Center (SPOC). Resources supporting this work were provided by the NASA High-End Computing (HEC) Program through the NASA Advanced Supercomputing (NAS) Division at Ames Research Center for the production of the SPOC data products. This research has made use of the Exoplanet Follow-up Observation Program website, which is operated by the California Institute of Technology, under contract with the National Aeronautics and Space Administration under the Exoplanet Exploration Program. This paper includes data collected by the TESS mission that are publicly available from the Mikulski Archive for Space Telescopes (MAST).

This article is based on observations made with the MuSCAT2 instrument, developed by ABC, at Telescopio Carlos S\'anchez operated on the island of Tenerife by the IAC in the Spanish Observatorio del Teide. 

This paper is based on observations made with the MuSCAT3 instrument, developed by the Astrobiology Center and under financial supports by JSPS KAKENHI (JP18H05439) and JST PRESTO (JPMJPR1775), at Faulkes Telescope North on Maui, HI, operated by the Las Cumbres Observatory.

Based on observations made with the Nordic Optical Telescope, operated by the Nordic Optical Telescope Scientific Association at the Observatorio del Roque de los Muchachos, La Palma, Spain, of the Instituto de Astrofisica de Canarias.  
The data presented here were obtained in part with ALFOSC, which is provided by the Instituto de Astrofisica de Andalucia (IAA) under a joint agreement with the University of Copenhagen and NOTSA.  

Based on observations obtained with the Samuel Oschin 48 inch Telescope at the Palomar Observatory as part of the Zwicky Transient Facility project. ZTF is supported by the National Science Foundation under grant No. AST-1440341 and a collaboration including Caltech, IPAC, the Weizmann Institute for Science, the Oskar Klein Center at Stockholm University, the University of Maryland, the University of Washington, Deutsches Elektronen-Synchrotron and Humboldt University, Los Alamos National Laboratories, the TANGO Consortium of Taiwan, the University of Wisconsin at Milwaukee, and Lawrence Berkeley National Laboratories. Operations are conducted by COO, IPAC, and UW.

This work is partly supported by JSPS KAKENHI grant Nos. JP17H04574, JP18H01265, and JP18H05439, Grant-in-Aid for JSPS Fellows grant No. JP20J21872, JST PRESTO grant No. JPMJPR1775, and a University Research Support Grant from the National Astronomical Observatory of Japan (NAOJ). 
This work is also partly financed by the Spanish Ministry of Economics and Competitiveness through grants PGC2018-098153-B-C31 and PID2019-109522GB-C53. 
J.K. gratefully acknowledges the support of the Swedish National Space Agency (DNR 2020-00104). G.M. has received funding from the European Union's Horizon 2020 research and innovation program under the Marie Sk\l{}odowska-Curie grant agreement No. 895525. M.T. is supported by JSPS KAKENHI grant Nos. 18H05442, 15H02063, and 22000005.

\end{acknowledgements}

\facilities{TESS, NOT (ALFOSC), Keck:II (NIRC2), Sanchez (MuSCAT2), FTN (Spectral, MuSCAT3), LCOGT (Sinistro), GTC (OSIRIS), MAST}

\software{{\tt PyAstronomy} \citep{pya}, {\tt emcee} \citep{2013PASP..125..306F}, {\tt celerite} \citep{2017AJ....154..220F}, {\tt PyTransit} \citep{Parviainen2015}, {\tt LDTk} \citep{Parviainen2015b}, {\tt TTVFast} \citep{2014ApJ...787..132D}, {\tt OpenTS} (\url{https://github.com/hpparvi/opents}), \texttt{SPOCK} \citep{spock_2020}}

\appendix
\section{\rev{Other transit observations}}

\rev{In this section we summarize ground-based transit observations of the TOI-1749 planets that were not used in the main analyses of this work.}

We observed one transit of TOI-1749c in $R$ band on 2020 April 10 UT with the 0.8m telescope at the George Mason University Observatory in Fairfax, Virginia, with an exposure time of 120~s. Unfortunately, we could not robustly extract a transit signal from this data set, likely because it was impacted by uncorrectable systematics originating from large tracking errors. We have decided not to use this data set for further analyses.

We also observed one transit of TOI-1749c in $R$ band on 2020 August 14 UT with ALFOSC on the 2.56~m NOT. We used an optical diffuser such that the FWHM of the stellar \rev{PSF} was $\sim$25 pixels ($\sim$5\arcsec) and an exposure time of 300~s. We do not use this data set for further analyses because of the low significance of the transit signal \footnote{We could not separate time-correlated noise from the transit signal due to the low sampling rate and a lack of long enough out-of-transit baseline.}.

In addition, we observed one transit of TOI-1749d in $I$ band on 2020 July 10 UT with the 0.4\,m telescope at the Observatori Astron\`{o}mic Albany\`{a} (OAA) near Albanya, Spain, with an exposure time of 140~s. We do not use this data set for further analyses because of a low significance of the transit signal.

Finally, we observed an expected transit of TOI-1749b on 2020 October 10 UT with OSIRIS on the 10m GTC at the Roque de los Muchachos Observatory, in an attempt to confirm the transit signal of this candidate. To check for an achromaticity of transit depth, we conducted spectrophotometric observation using the R1000R grism and a long slit with a width of 40$''$, which was aligned to the target and two comparison stars with a position angle of 211.45$^\circ$. The observation was done between 21:18 UT (airmass = 1.4) and 00:45 UT (airmass = 2.3) with an exposure time of 13~s. Unfortunately, the obtained data were heavily affected by systematics of instrumental origin as well as by thin cirrus crossing the sky after 23:30 UT. Despite exhaustive efforts, we could not correct the systematics to the level that is required to detect the expected transit signal with a depth of only 0.05\%. We have therefore decided not to use this data set for further analyses.

We note that, in these observations, we did not see any evidence of false positives for the planetary candidates due to contamination from eclipsing binaries (which could exhibit deeper transits on the target or neighboring stars).

\bibliography{ref}{}
\bibliographystyle{aasjournal}



\end{document}